\documentclass[
	groupedaddress,
	amsmath,amssymb,
	prb,
	floatfix,
	superscriptaddress,
    twocolumn,
	aps
 ]{revtex4-2}
 
\usepackage{graphicx} 
\usepackage{amsmath}
\usepackage{xcolor}
\usepackage{hyperref}
\usepackage{cleveref}
\usepackage[english]{babel}

\newcommand{\Va}{\mathrm{Va}}

\begin{document}

\title{First-principles calculations of the thermodynamic factor of multicomponent alloys}
\author{Damien K. J. Lee}
\affiliation{Laboratory of materials design and simulation (MADES), Institute of Materials, \'{E}cole Polytechnique F\'{e}d\'{e}rale de Lausanne}
\author{Shashank Saxena}
\affiliation{Laboratory of materials design and simulation (MADES), Institute of Materials, \'{E}cole Polytechnique F\'{e}d\'{e}rale de Lausanne}
\author{Anton Van der Ven}
\affiliation{Materials department, University of California, Santa Barbara}
\author{Anirudh Raju Natarajan}
\email{anirudh.natarajan@epfl.ch}
\affiliation{Laboratory of materials design and simulation (MADES), Institute of Materials, \'{E}cole Polytechnique F\'{e}d\'{e}rale de Lausanne}

\begin{abstract}
Computing non-dilute diffusion coefficients from Onsager transport coefficients of a multicomponent alloy requires the thermodynamic factor, which measures the curvature of the free energy with respect to composition. As diffusion in metals is mediated by vacancies, the curvature that matters is not that of the vacancy-free alloy, which is straightforward to compute, but that of an alloy carrying a dilute concentration of vacancies. The resulting matrix is nearly singular, and composition fluctuations from which it is obtained converge slowly in Monte Carlo simulations. Here we develop two routes that avoid sampling vacancy composition fluctuations. The first constructs the thermodynamic factor from two quantities, the free-energy curvature of the vacancy-free alloy and the vacancy concentration. The second truncates the semi-grand canonical partition function at a single vacancy and recovers the full thermodynamic factor of the vacancy-containing alloy. The same separation shows that the largest eigenvalue of the diffusion matrix is the vacancy tracer diffusion coefficient for any number of components and any degree of non-ideality, while the remaining eigenvalues scale with the vacancy concentration. Across binary through quinary alloys of the Hf--Mo--Nb--Ti--Zr system, both routes reproduce eigenvalues and eigenvectors of diffusion matrices obtained from fully converged Monte Carlo simulations. The single-vacancy expansion reaches this accuracy with roughly one tenth of the sampling effort. A scaling relation extends both routes to local vacancy chemical potentials away from equilibrium, without additional simulations. These results allow thermodynamic descriptions from atomistic models or CALPHAD assessments to be used directly in mesoscale simulations of mass transport.
\end{abstract}

\maketitle

\section{Introduction}

Substitutional diffusion in multicomponent alloys is governed by the interplay between kinetic transport coefficients and thermodynamic driving forces~\cite{darken_formal_1951,allnatt_atomic_2003,balluffi_kinetics_2005,mishin_atomistic_1997,mishin_atomistic_1997-1}. The matrix of diffusion coefficients $\mathbf{D}$ relates the flux of each species to gradients in composition. It is the product of the Onsager transport coefficient matrix $\mathbf{\tilde{L}}$ and the thermodynamic factor $\mathbf{\tilde{\Theta}}$, where $\mathbf{D}=\mathbf{\tilde{L}}\mathbf{\tilde{\Theta}}$~\cite{kehr_mobility_1989,van_der_ven_vacancy_2010}. The thermodynamic factor is proportional to the Hessian of the Gibbs free energy with respect to composition and encodes how the exchange chemical potentials \cite{cahn_invariant_1983} vary with alloy composition. The sign of its eigenvalues determines whether a phase is stable against compositional fluctuations. All three matrices depend on the alloy composition, the vacancy concentration, and the temperature, so mesoscale simulations of mass transport must recompute them at every time step from the local composition of each region held in local equilibrium. For metallic alloys, where diffusion is vacancy-mediated, $\mathbf{\tilde{\Theta}}$ must be evaluated in the dilute-vacancy limit. The low vacancy concentration renders the matrix nearly singular and difficult to compute accurately.

Multi-principal element alloys (MPEAs) contain several principal elements rather than a single base element, and display a wide range of high-temperature, mechanical, and functional properties~\cite{yeh_nanostructured_2004,george_high-entropy_2019,miracle_critical_2017}. Many form single-phase disordered solid solutions on simple crystal structures, with properties that depend on composition, temperature, and the degree of chemical short-range order. Predicting microstructural evolution and high-temperature performance in MPEAs therefore requires accurate diffusion coefficients across the multicomponent composition space. Diffusion in MPEAs is often assumed to be sluggish, but the evidence is mixed. Some experiments report reduced diffusivities~\cite{tsai_sluggish_2013}, whereas first-principles calculations reveal composition-dependent behavior that can be either sluggish or anti-sluggish~\cite{lee_diffusion_2026,daw_sluggish_2021,sen_anti-sluggish_2023}. Resolving this question requires methods to compute both $\mathbf{\tilde{L}}$ and $\mathbf{\tilde{\Theta}}$ from atomistic models.

The thermodynamic factor can be obtained from free energy databases assessed within the CALPHAD formalism or from atomistic simulations informed by a surrogate energy model \cite{gao_thermodynamics_2021,kattner_need_2020,wu_co-based_2022}. CALPHAD assessments of substitutional alloys do not generally treat the vacancy as an explicit species, so the second derivatives they provide describe the vacancy-free alloy rather than the dilute-vacancy limit that $\mathbf{D}$ requires. Atomistic simulations that sample the vacancy alongside the chemical species give the required derivatives directly, but at a sampling cost that grows as the vacancy concentration becomes dilute \cite{belak_effect_2015,goiri_role_2019,lee_modeling_2026}.

Cluster expansions parameterize the configurational energy and migration barriers of chemical decorations of several elements on an underlying parent crystal structure~\cite{sanchez_generalized_1984,van_der_ven_first-principles_2001,natarajan_machine-learning_2018}. They map energies from computationally expensive first-principles electronic structure calculations onto a faster surrogate model that can be sampled with Monte Carlo methods to evaluate finite-temperature thermodynamic properties. The embedded cluster expansion (eCE) extends this approach to alloys with several components using chemical embeddings that reduce the number of independent parameters~\cite{muller_constructing_2025}. Combined with Monte Carlo simulations, the eCE has been used to compute equilibrium vacancy concentrations in MPEAs as a function of composition and temperature~\cite{lee_modeling_2026}, and coupled with kinetic Monte Carlo to evaluate the full matrix of non-dilute diffusion coefficients from first principles~\cite{lee_diffusion_2026}. Evaluating the thermodynamic factor in the dilute-vacancy limit nevertheless remains a bottleneck in these calculations.

In this work we develop two routes to the thermodynamic factor of a multicomponent alloy in the dilute-vacancy limit. The first approximates $\mathbf{\tilde{\Theta}}$ from the covariance matrix of composition fluctuations in the vacancy-free alloy together with the equilibrium vacancy concentration, so that vacancies never enter the sampled ensemble. The second extends an existing sampling framework into a single-vacancy expansion of the covariance matrix with vacancies treated explicitly, serving both as an independent method and as a benchmark for the first. We compare both routes against the inverse of converged covariance matrices from long semi-grand canonical Monte Carlo simulations of an alloy containing vacancies, and quantify the effect of each approximation on the eigenspectrum of the diffusion matrix for binary, ternary, quaternary, and quinary alloys. We then derive a rescaling relation that extends these results to vacancy chemical potentials away from equilibrium.

\section{Formalism}
\label{sec:formalism}

Flux expressions for a substitutional solid relate the fluxes of the $c$ chemical elements and of the vacancies, $[J_{1},J_{2},\cdots,J_{c},J_{\mathrm{Va}}]^\mathrm{T}$, to spatial gradients in the chemical potentials, $[\nabla \mu_{1},\nabla \mu_{2},\cdots,\nabla \mu_{c},\nabla \mu_{\mathrm{Va}}]^\mathrm{T}$. Within a region of the crystal that conserves its total number of lattice sites, the $c+1$ fluxes must sum to zero. Combined with the Onsager reciprocity relations, this constraint leaves $c$ independent fluxes of the chemical elements, $\mathbf{J} = [J_{1},J_{2},\cdots,J_{c}]^\mathrm{T}$, driven by gradients in the exchange chemical potentials $\tilde{\mu}_{i} = \mu_{i} - \mu_{\mathrm{Va}}$:
\begin{equation}
  \label{eq:onsager_flux}
  \mathbf{J} = -\mathbf{L} \nabla \boldsymbol{\tilde{\mu}}
\end{equation}
where the matrix $\mathbf{L}$ holds the Onsager transport coefficients. Mesoscale simulations of mass transport track compositions rather than chemical potentials, so it is convenient to express the fluxes in terms of composition gradients:
\begin{equation}
  \label{eq:fickian_flux}
  \mathbf{J} = - \mathbf{\tilde{L}} \mathbf{\tilde{\Theta}} \nabla \mathbf{C}
\end{equation}
where the elements of $\mathbf{C} = [C_{1},C_{2},\cdots,C_{c}]^\mathrm{T}$ are the concentrations $C_{i} = x_{i}/\Omega$, the number of atoms of species $i$ per unit volume, with $\Omega$ the volume per substitutional site. Absorbing a factor of $k_\mathrm{B}T\Omega$ into the transport coefficients defines $\mathbf{\tilde{L}} = (k_\mathrm{B} T \Omega)\mathbf{L}$. Converting the gradients in exchange chemical potential to gradients in composition requires the derivatives that make up the thermodynamic factor $\mathbf{\tilde{\Theta}}$:
\begin{equation}
	\label{eq:thermodynamic_factor}
	\mathbf{\tilde{\Theta}}= \frac{1}{k_\mathrm{B}T}
	\begin{bmatrix}
		\frac{\partial\tilde{\mu}_1}{\partial x_1} & \frac{\partial\tilde{\mu}_1}{\partial x_2} & \dots  & \frac{\partial\tilde{\mu}_1}{\partial x_c} \\
		\frac{\partial\tilde{\mu}_2}{\partial x_1} & \frac{\partial\tilde{\mu}_2}{\partial x_2} & \dots  & \frac{\partial\tilde{\mu}_2}{\partial x_c} \\
		\vdots                                     & \vdots                                     & \ddots & \vdots                                     \\
		\frac{\partial\tilde{\mu}_c}{\partial x_1} & \frac{\partial\tilde{\mu}_c}{\partial x_2} & \dots  & \frac{\partial\tilde{\mu}_c}{\partial x_c}
	\end{bmatrix}
\end{equation}

The product $\mathbf{\tilde{L}}\mathbf{\tilde{\Theta}}$ in \cref{eq:fickian_flux} is the matrix of diffusion coefficients $\mathbf{D}$. Its two factors are not equally easy to compute. Kinetic Monte Carlo simulations informed by an atomistic surrogate model such as a cluster expansion \cite{goiri_role_2019} give $\mathbf{\tilde{L}}$ directly, whereas the thermodynamic factor of \cref{eq:thermodynamic_factor} must be evaluated at the vanishingly small vacancy concentrations that mediate diffusion in a metallic alloy. This section derives two methods for computing $\mathbf{\tilde{\Theta}}$ from atomistic simulation. Using a prototypical binary alloy, we first show that a coordinate transformation of the compositions, fluxes and chemical potentials separates $\mathbf{\tilde{\Theta}}$ and the eigenspectrum of $\mathbf{D}$ into a contribution set by vacancy thermodynamics and one set entirely by the thermodynamics of the vacancy-free alloy. We then generalize this separation to an arbitrary number of components, derive an alternative approximation to $\mathbf{\tilde{\Theta}}$ from the semi-grand canonical partition function truncated at a single vacancy, and benchmark both routes against converged Monte Carlo simulations of alloys in the Hf--Mo--Nb--Ti--Zr system.

\subsection{Coordinate transforms of fluxes, compositions and chemical potentials in a binary alloy}
\label{sec:coord-transf-flux}

\begin{figure}[h!]
  \centering
  \includegraphics[width=0.29\textwidth]{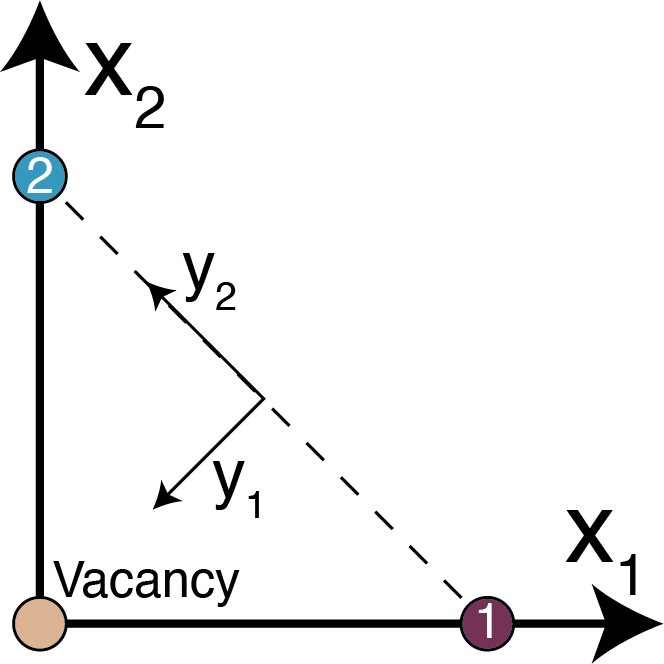}
  \caption{Composition axes of a prototypical binary substitutional alloy. The elemental compositions $x_{1}$ and $x_{2}$ are measured from the vacancy corner at the origin, and the dashed line joining the two pure elements marks the vacancy-free alloy, $x_{1}+x_{2}=1$. The transformed coordinates of \cref{eq:composition_transformation}, $y_{1}$ and $y_{2}$, are directed perpendicular and parallel to this line, so that $y_{1}$ tracks the vacancy concentration and $y_{2}$ the degree of intermixing between the two elements.}
  \label{fig:binary_composition_schematic}
\end{figure}

The thermodynamic factor of \cref{eq:thermodynamic_factor} is defined with respect to a particular set of composition variables. Choosing those variables well makes the vacancy contribution to $\mathbf{\tilde{\Theta}}$ appear as a single divergent term and leaves a remainder that belongs entirely to the vacancy-free alloy. We work out the binary alloy in full, since its diffusion matrix is small enough to diagonalize analytically and the structure that emerges carries over to the multicomponent generalization.

Consider a binary alloy in which atoms of species 1 and 2 and vacancies are distributed over $M$ substitutional sites, with compositions $x_{1} = N_{1}/M$, $x_{2} = N_{2}/M$ and $x_{\mathrm{Va}} = N_{\mathrm{Va}}/M$. The elemental compositions and the corresponding concentrations make up the vectors $\mathbf{x} = [x_{1},x_{2}]^{\mathrm{T}}$ and $\mathbf{C} = \mathbf{x}/\Omega$, while the fluxes of the two elements and their conjugate exchange chemical potentials are the $\mathbf{J}_{\mathbf{x}}$ and $\boldsymbol{\tilde{\mu}}_{\mathbf{x}}$ of \cref{eq:onsager_flux,eq:fickian_flux}. The subscript denotes the compositional basis within which both quantities are measured.

The composition axes defined by $\mathbf{x}$ are not the most convenient basis for describing vacancy-mediated diffusion. In the dilute vacancy regime, diffusion among several elements and vacancies is conventionally partitioned into vacancy transport and interdiffusion among the chemical species \cite{darken_diffusion_1948,kehr_mobility_1989,van_der_ven_vacancy_2010}. As \cref{fig:binary_composition_schematic} shows, a displacement along either axis of $\mathbf{x}$ changes the vacancy concentration and the degree of intermixing at the same time, so vacancy transport and interdiffusion appear together in every element of the flux expressions. We therefore move to a basis in which one coordinate tracks the vacancy concentration and the other tracks the intermixing between the two elements. A linear transformation of the compositions is:
\begin{equation}
  \label{eq:composition_transformation}
  \mathbf{y} = \mathbf{Q} \mathbf{x} + \mathbf{y}_{0}
\end{equation}
where $\mathbf{Q}$ is a square invertible matrix and $\mathbf{y}_{0}$ shifts the origin. Any invertible $\mathbf{Q}$ and origin shift are admissible. Without loss of generality, for the binary alloy we adopt:
\begin{align}
  \mathbf{Q} &= \frac{1}{2}
  \begin{bmatrix}
    -1 & -1 \\
    -1 & 1
  \end{bmatrix}
  \label{eq:binary_linear_transformation} \\
  \mathbf{y}_{0} &= \frac{1}{2}
  \begin{bmatrix}
    1 \\
    1
  \end{bmatrix}
  \label{eq:binary_origin_shift}
\end{align}

The resulting coordinates $\mathbf{y} = [y_{1},y_{2}]^{\mathrm{T}}$ are $y_{1} = x_{\mathrm{Va}}/2$ and $y_{2} = (1-x_{1}+x_{2})/2$, shown in \cref{fig:binary_composition_schematic}. The first is proportional to the vacancy concentration, and the second reduces to $x_{2}$ along the vacancy-free line; thus, $y_{1}$ and $y_{2}$ measure displacements perpendicular and parallel to that line. Fluxes of the individual chemical species transform like the elemental compositions, $\mathbf{J}_{\mathbf{y}} = \mathbf{Q} \mathbf{J}_{\mathbf{x}}$. The chemical potentials are conjugate to the compositions, so they must transform with the inverse transpose, $\boldsymbol{\tilde{\mu}}_{\mathbf{y}} = \mathbf{Q}^{-\mathrm{T}}\boldsymbol{\tilde{\mu}}_{\mathbf{x}}$. This ensures that the rate of entropy production $(\dot{\sigma})$ is invariant under the change of basis:
\begin{equation}
  \label{eq:entropy_generation_invariance}
  -T\dot{\sigma} = \mathbf{J}_{\mathbf{x}}^{\mathrm{T}} \nabla \boldsymbol{\tilde{\mu}}_{\mathbf{x}} = \mathbf{J}_{\mathbf{x}}^{\mathrm{T}} \mathbf{Q}^{\mathrm{T}} \mathbf{Q}^{-\mathrm{T}} \nabla \boldsymbol{\tilde{\mu}}_{\mathbf{x}} = \mathbf{J}_{\mathbf{y}}^{\mathrm{T}} \nabla \boldsymbol{\tilde{\mu}}_{\mathbf{y}}
\end{equation}

Because the fluxes and the chemical potentials transform in these ways, the matrices that connect them are constrained by the transformation:
\begin{align}
  \mathbf{\tilde{L}}_{\mathbf{y}} &= \mathbf{Q} \mathbf{\tilde{L}}_{\mathbf{x}} \mathbf{Q}^{\mathrm{T}}
  \label{eq:mobility_transformation} \\
  \mathbf{\tilde{\Theta}}_{\mathbf{y}} &= \mathbf{Q}^{-\mathrm{T}} \mathbf{\tilde{\Theta}}_{\mathbf{x}} \mathbf{Q}^{-1}
  \label{eq:thermodynamic_factor_transformation} \\
  \mathbf{D}_{\mathbf{y}} &= \mathbf{Q} \mathbf{D}_{\mathbf{x}} \mathbf{Q}^{-1}
  \label{eq:diffusion_coefficient_transformation}
\end{align}
\noindent $\mathbf{\tilde{L}}$ and $\mathbf{\tilde{\Theta}}$ undergo congruence transformations, effected by $\mathbf{Q}^{\mathrm{T}}$ and $\mathbf{Q}^{-1}$ respectively, whereas $\mathbf{D}$ undergoes a similarity transformation and therefore has the same eigenvalues in either basis. The transformation changes nothing physical, only the axes along which the fluxes and the curvatures of the free energy are resolved.

Both matrices, $\mathbf{\tilde{\Theta}}$ and $\mathbf{\tilde{L}}$, take on a simpler form in the new basis in the limit of dilute vacancy concentrations. Consider the Onsager transport coefficients first, obtained from \cref{eq:mobility_transformation}:
\begin{equation}
  \label{eq:transformed_mobility}
  \mathbf{\tilde{L}}_{\mathbf{y}} =  \frac{1}{4}
  \begin{bmatrix}
    D^{\star}_{\mathrm{Va}}x_{\mathrm{Va}} & \tilde{L}_{11}-\tilde{L}_{22} \\
    \tilde{L}_{11}-\tilde{L}_{22} & \tilde{L}_{11}+\tilde{L}_{22}-2\tilde{L}_{12}
  \end{bmatrix}
\end{equation}
\noindent where $\tilde{L}_{ij}$ are the elements of $\mathbf{\tilde{L}}_{\mathbf{x}}$ and the $(1,1)$ element has been rewritten in terms of the tracer diffusion coefficient of the vacancy using $\tilde{L}_{11}+\tilde{L}_{22}+2\tilde{L}_{12} = D^{\star}_{\mathrm{Va}}x_{\mathrm{Va}}$. 
In the dilute vacancy limit and in the absence of vacancy cluster diffusion mechanisms \cite{van_der_ven_understanding_2012,kolli_elucidating_2021}, the elements of $\mathbf{\tilde{L}}_{\mathbf{x}}$ are proportional to the vacancy concentration $x_{\mathrm{Va}}$ \cite{kehr_mobility_1989,van_der_ven_vacancy_2010}.
The flux, $J_{y_1} = - J_{x_2}-J_{x_1}$ in the new coordinate system is proportional to the vacancy flux, so this element relates the vacancy flux to the gradient in the vacancy chemical potential. Similarly, the interdiffusion flux is coupled to the gradient in $y_{2}$ through $\tilde{L}_{11}+\tilde{L}_{22}-2\tilde{L}_{12}$.  As with $\mathbf{\tilde{L}}_{\mathbf{x}}$, every element of $\mathbf{\tilde{L}}_{\mathbf{y}}$ is proportional to $x_{\mathrm{Va}}$. 

\begin{figure}
	\centering
	\includegraphics[width=0.4\textwidth]{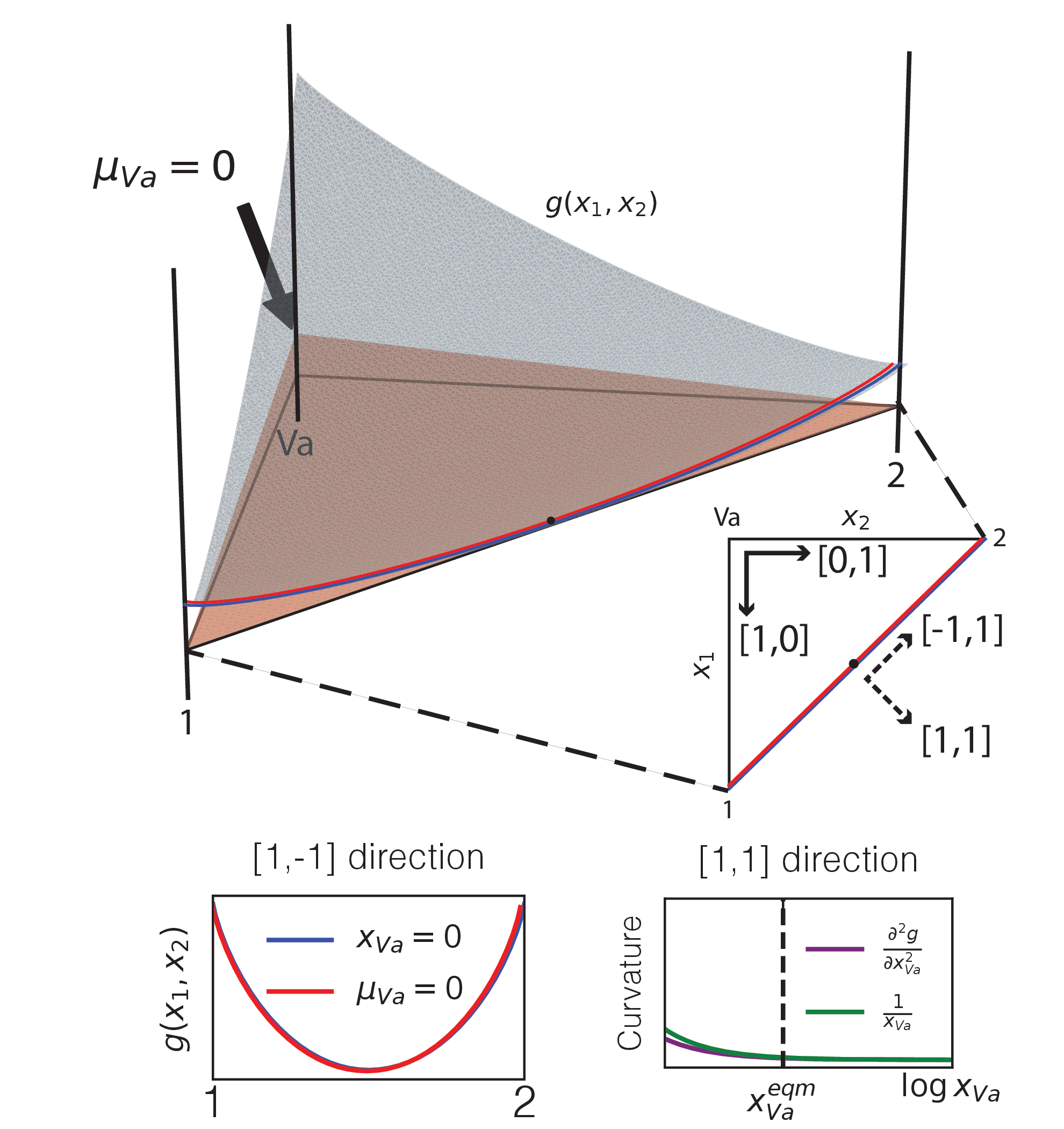}
	\caption{Free energy surface of a binary alloy of species A and B containing vacancies described by an ideal solution model.}
	\label{fig:free_energy}
\end{figure}

The same transformation reveals structure in the thermodynamic factor. The chemical potential of any species, including the vacancy, can be written as $\mu_{i} = \mu_{i}^{0}+k_\mathrm{B}T\log \left(\gamma_{i}x_{i}\right)$, where the activity coefficient $\gamma_{i}$ measures the deviation from ideality and generally depends on alloy composition. For a dilute species this expression reduces to the Henrian form. Substituting these chemical potentials into \cref{eq:thermodynamic_factor} gives the thermodynamic factor of the binary alloy:
\begin{equation}
	\mathbf{\tilde{\Theta}}_{\mathbf{x}} =
  \begin{bmatrix}
    \frac{1}{x_{\mathrm{Va}}}+\frac{1}{x_1} + f_{11} & \frac{1}{x_{\mathrm{Va}}}+f_{12} \\ \frac{1}{x_{\mathrm{Va}}}+f_{21} & \frac{1}{x_{\mathrm{Va}}}+\frac{1}{x_2} + f_{22}
  \end{bmatrix}
  \label{eq:binary_thermodynamic_factor_matrix_activity}
\end{equation}
\noindent where $f_{ij} = \partial \log{\gamma_{i}}/\partial x_{j} - \partial \log{\gamma_{\mathrm{Va}}}/\partial x_{j}$. Every element of \cref{eq:binary_thermodynamic_factor_matrix_activity} diverges as the vacancies become dilute, and each carries information about the vacancy both directly and through the vacancy activity coefficient in $f_{ij}$. These elements are the curvatures of the Gibbs free energy along the elemental composition axes, illustrated schematically in \cref{fig:free_energy}.

Measuring the curvatures along the $\mathbf{y}$ axes rather than the $\mathbf{x}$ axes separates the two contributions. The transformed thermodynamic factor is:
\begin{equation}
  \label{eq:transformed_binary_thermodynamic_factor_compact}
  \mathbf{\tilde{\Theta}}_{\mathbf{y}} =
    \begin{bmatrix}
    \frac{4}{x_{\mathrm{Va}}}+a & b\\ b & \Theta_{\mathrm{mix}}
  \end{bmatrix}
\end{equation}
\noindent whose three distinct elements are the curvature perpendicular to the vacancy-free line, $\partial^{2}g/\partial y_{1}^{2}$, the curvature along it, $\partial^{2}g/\partial y_{2}^{2} = \Theta_{\mathrm{mix}}$, and the coupling between the two directions, $\partial^{2}g/\partial y_{1}\partial y_{2} = b$. The coefficient $a$ is what remains of $\partial^{2}g/\partial y_{1}^{2}$ once the divergent term is separated out. \Cref{sec-app:binary-alloy} gives the full matrix and the definitions of $a$ and $b$ in \cref{eq:theta_y_a,eq:theta_y_b}. The curvature along the vacancy-free line is
\begin{equation}
  \label{eq:theta_mix}
  \Theta_{\mathrm{mix}} = \frac{1}{x_{1}}+\frac{1}{x_{2}} + \frac{\partial \log{\gamma_{1}}}{\partial x_{1}} + \frac{\partial \log{\gamma_{2}}}{\partial x_{2}} - \frac{\partial \log{\gamma_{1}}}{\partial x_{2}} - \frac{\partial \log{\gamma_{2}}}{\partial x_{1}}
\end{equation}

\noindent Written this way, the entire $x_{\mathrm{Va}}$ dependence of $\mathbf{\tilde{\Theta}}_{\mathbf{y}}$ is confined to the single term $4/x_{\mathrm{Va}}$, and none of $a$, $b$ or $\Theta_{\mathrm{mix}}$ contains $x_{\mathrm{Va}}$.

The three quantities ($\Theta_{\mathrm{mix}},a,b$) differ in how much of the vacancy thermodynamics they retain. The coefficients $a$ and $b$ both contain derivatives of $\gamma_{\mathrm{Va}}$, so evaluating them requires knowing how the vacancy interacts with the alloy and how that interaction changes with alloy composition. By contrast, the derivatives of $\gamma_{\mathrm{Va}}$ cancel identically in \cref{eq:theta_mix}, leaving $\Theta_{\mathrm{mix}}$ dependent only on the activity coefficients of species 1 and 2. What remains is the curvature of the vacancy-free alloy along its own composition axis. The insets of \cref{fig:free_energy} illustrate this for an ideal free energy model. The curvature along $y_{2}$ is nearly indistinguishable from that of the vacancy-free alloy, whereas the curvature along $y_{1}$ grows as $1/x_{\mathrm{Va}}$.

The eigenvalues and eigenvectors of the diffusion matrix, $\mathbf{D} = \mathbf{\tilde{L}}\mathbf{\tilde{\Theta}}$, set the rates of vacancy diffusion and of interdiffusion \cite{kehr_mobility_1989,van_der_ven_vacancy_2010}. Multiplying \cref{eq:transformed_mobility} by \cref{eq:transformed_binary_thermodynamic_factor_compact} gives the diffusion matrix of the binary alloy in the transformed coordinates, listed in full as \cref{eq:transformed_diffusion_matrix}. Because $\mathbf{D}_{\mathbf{y}}$ is a $2\times 2$ matrix, it can be diagonalized analytically. Following the order of operations outlined by \citeauthor{van_der_ven_vacancy_2010}~\cite{van_der_ven_vacancy_2010} and collecting the resulting terms according to their power of the vacancy concentration gives the eigenvalues of the binary alloy:
\begin{align}
  \lambda_{+} &= D^{\star}_{\mathrm{Va}} + \mathcal{O}(x_{\mathrm{Va}})
  \label{eq:eigenvalue_binary_plus} \\
  \lambda_{-} &= \frac{\tilde{L}_{11}\tilde{L}_{22}-\tilde{L}_{12}^{2}}{x_{\mathrm{Va}}D^{\star}_{\mathrm{Va}}}\Theta_{\mathrm{mix}} + \mathcal{O}(x_{\mathrm{Va}}^{2})
  \label{eq:eigenvalue_binary_minus}
\end{align}
\noindent Every Onsager coefficient is itself proportional to the vacancy concentration, so the leading term of \cref{eq:eigenvalue_binary_minus} is of order $x_{\mathrm{Va}}$ and its remainder of order $x_{\mathrm{Va}}^{2}$. The two eigenvalues of \cref{eq:eigenvalue_binary_minus,eq:eigenvalue_binary_plus} behave differently as vacancies become dilute. Vacancy diffusion proceeds at the finite rate $D^{\star}_{\mathrm{Va}}$, whereas interdiffusion slows in proportion to the vacancy concentration, separating the two eigenvalues by several orders of magnitude in a typical metallic alloy. The thermodynamic factor enters the eigenvalues only through $\Theta_{\mathrm{mix}}$ and $x_{\mathrm{Va}}$ in $\lambda_{-}$.

The corresponding eigenvectors in the $\mathbf{y}$ basis are derived in \cref{sec-app:binary-alloy}. As $x_{\mathrm{Va}}\rightarrow 0$, the $(1,2)$ element of $\mathbf{D}_{\mathbf{y}}$ vanishes while the $(2,1)$ element remains finite, and the eigenvectors reduce to
\begin{align}
  \mathbf{v}_{+} &\rightarrow \left[\,D^{\star}_{\mathrm{Va}},\ \left(\tilde{L}_{11}-\tilde{L}_{22}\right)/x_{\mathrm{Va}}\,\right]^{\mathrm{T}}
  \label{eq:eigenvector_binary_plus} \\
  \mathbf{v}_{-} &\rightarrow \left[\,0,\ 1\,\right]^{\mathrm{T}}
  \label{eq:eigenvector_binary_minus}
\end{align}
\noindent The slow mode aligns with the $y_{2}$ axis and therefore describes interdiffusion at fixed vacancy concentration. The fast mode is the vacancy mode and carries an admixture of interdiffusion set by the difference in transport coefficients $\tilde{L}_{11}-\tilde{L}_{22}$.

\Cref{eq:eigenvalue_binary_plus,eq:eigenvalue_binary_minus,eq:eigenvector_binary_plus,eq:eigenvector_binary_minus} show that neither $a$ nor $b$ survives in the dilute-vacancy limit. The two coefficients that retain the vacancy activity coefficient are therefore the two that the eigenspectrum does not need. In \cref{eq:transformed_diffusion_matrix}, $a$ appears only in products that already carry a factor of the vacancy concentration, and $b$ appears only in combinations that contribute at order $x_{\mathrm{Va}}$ to \cref{eq:eigenvalue_binary_plus} and at order $x_{\mathrm{Va}}^{2}$ to \cref{eq:eigenvalue_binary_minus}. Setting both to zero leaves the eigenvalues and eigenvectors of $\mathbf{D}$ unchanged to leading order in the vacancy concentration. Accurate values of $a$ and $b$ are needed to reproduce the thermodynamic factor of an alloy that holds vacancies, but not to compute its diffusion coefficients. The vacancy concentration and the curvature of the vacancy-free alloy suffice, and both are accessible without sampling an alloy and its vacancies in the same simulation.

\subsection{Approximating the thermodynamic factor in a multicomponent alloy}
\label{sec:appr-therm-fact}

A thermodynamic factor whose only vacancy dependence is the vacancy concentration itself, with every remaining contribution supplied by the vacancy-free alloy, can be constructed for a multicomponent alloy by following the strategy of \cref{sec:coord-transf-flux}. We first transform the fluxes, compositions and chemical potentials to a basis in which one coordinate is proportional to the vacancy concentration and the remaining coordinates leave that concentration unchanged, tracking instead the amount of each element in the alloy. The thermodynamic factor in this basis partitions in the same way as \cref{eq:transformed_binary_thermodynamic_factor_compact}. We then approximate it by setting the off-diagonal elements to zero, retaining only the term proportional to $1/x_{\mathrm{Va}}$ in the element that carries the vacancy concentration, and writing the remaining elements as curvatures of the vacancy-free alloy. Combining the result with an Onsager transport coefficient matrix obtained from kinetic Monte Carlo simulations gives the eigenspectrum of the diffusion matrix, accurate to corrections that vanish as the vacancy concentration goes to zero.

\begin{figure}[h!]
  \centering
  \includegraphics[width=0.45\textwidth]{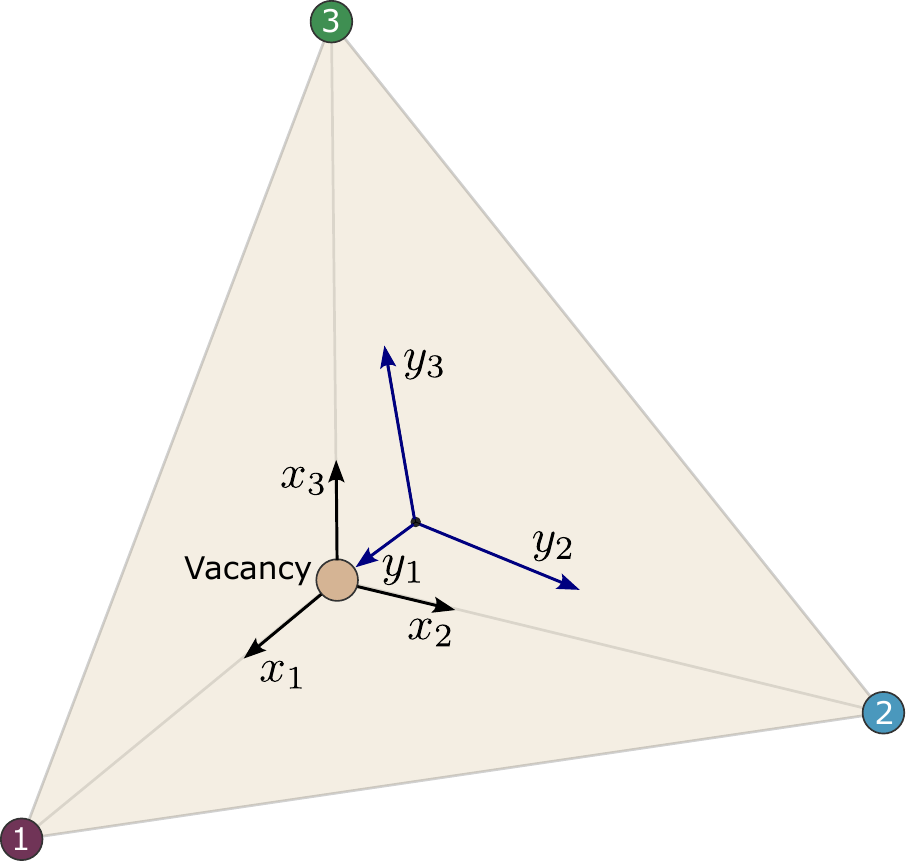}
  \caption{
  Composition axes of a ternary substitutional alloy. The elemental compositions $x_{1}$, $x_2$ and $x_{3}$ are measured from the vacancy corner at the origin, while the transformed coordinates $y_{1}$, $y_2$ and $y_{3}$ (see eq.~\eqref{eq:composition_transformation}) point towards the vacancy, element 2 and 3 from the centroid of the (1,1,1) plane.}
  \label{fig:axes_transformation}
\end{figure}

The coordinate transformation introduced for the binary alloy in \cref{eq:binary_linear_transformation,eq:binary_origin_shift} extends to any number of components. For an alloy of $c$ chemical elements, define
\begin{equation}
  \label{eq:general_composition_coordinate_transform}
  \mathbf{Q} = \frac{1}{c}
  \begin{bmatrix}
    -1     & -1     & \cdots & -1     \\
    -1     & c-1    & \cdots & -1     \\
    \vdots & \vdots & \ddots & \vdots \\
    -1     & -1     & \cdots & c-1
  \end{bmatrix}
  \qquad
  \mathbf{y}_{0} = \frac{1}{c}
  \begin{bmatrix}
    1      \\
    1      \\
    \vdots \\
    1
  \end{bmatrix}
\end{equation}

\noindent which reduces to \cref{eq:binary_linear_transformation,eq:binary_origin_shift} when $c=2$ and can be utilized in \cref{eq:composition_transformation}. Carrying out the multiplication gives $y_{1} = x_{\mathrm{Va}}/c$ and $y_{k} = x_{k} + x_{\mathrm{Va}}/c$ for $k \geq 2$, with $x_{1} = 1 - y_{1} - \sum_{k=2}^{c}y_{k}$ the dependent composition. The first coordinate $y_1$ is proportional to the vacancy concentration. Because $x_{\mathrm{Va}} = c\,y_{1}$ depends on $y_{1}$ alone, the remaining $c-1$ coordinates leave the vacancy concentration unchanged, and changes in concentration along them run parallel to the vacancy-free hyperplane $\sum_{i}x_{i}=1$. The hyperplane is the generalization of the vacancy-free line of \cref{fig:binary_composition_schematic}. The compositional coordinate transformation is schematically shown for a ternary alloy in \cref{fig:axes_transformation}.

The shift by $x_{\mathrm{Va}}/c$ that each $y_{k}$ carries has a geometric meaning. Adding $x_{\mathrm{Va}}/c$ to every elemental composition fills the vacant sites with an equal distribution of all $c$ elements, so the vacancies are removed exactly. The coordinates $y_{2},y_{3},\cdots,y_{c}$ are therefore the elemental compositions of the vacancy-free alloy reached by projecting the composition of the vacancy-bearing alloy onto the hyperplane $\sum_{i}x_{i}=1$ along the $[1,1,\cdots,1]^{\mathrm{T}}$ direction. As $x_{\mathrm{Va}}\rightarrow 0$ the shift vanishes and each $y_{k}$ reduces to $x_{k}$, so in the dilute-vacancy limit the transformed coordinates are simply the elemental compositions of the vacancy-free alloy. Any set of coordinates $y_{2},\cdots,y_{c}$ serves equally well, provided they form a complete basis and remain orthogonal to $y_{1}$, and hence to the vacancy concentration.

The inverse of \cref{eq:general_composition_coordinate_transform} is $\mathbf{Q}^{-1} = \begin{bmatrix} -\vec{1} & \hat{\mathbf{Q}} \end{bmatrix}$, where $\hat{\mathbf{Q}}$ is a $c\times (c-1)$ matrix:
\begin{equation}
  \label{eq:composition_projection_matrix}
  \hat{\mathbf{Q}} =
  \begin{bmatrix}
    -1     & -1     & \cdots & -1     \\
    1      & 0      & \cdots & 0      \\
    0      & 1      & \cdots & 0      \\
    \vdots & \vdots & \ddots & \vdots \\
    0      & 0      & \cdots & 1
  \end{bmatrix}
\end{equation}
and $\vec{1}$ is the $c$-dimensional vector of ones $\vec{1} = [1, 1, \cdots, 1]^{\mathrm{T}}$. The block form of $\mathbf{Q}^{-1}$ makes the coordinate transform of the thermodynamic factor $\mathbf{\tilde{\Theta}}$ straightforward to compute:
\begin{equation}
  \label{eq:general_transformed_thermodynamic_factor}
  \tilde{\mathbf{\Theta}}_{\mathbf{y}} =
  \begin{bmatrix}
    \vec{1}^{\,\mathrm{T}} \tilde{\mathbf{\Theta}}_{\mathbf{x}} \vec{1}      &
    -\vec{1}^{\,\mathrm{T}} \tilde{\mathbf{\Theta}}_{\mathbf{x}} \hat{\mathbf{Q}}          \\
    -\hat{\mathbf{Q}}^\mathrm{T} \tilde{\mathbf{\Theta}}_{\mathbf{x}} \vec{1} &
    \hat{\mathbf{Q}}^\mathrm{T} \tilde{\mathbf{\Theta}}_{\mathbf{x}} \hat{\mathbf{Q}}
  \end{bmatrix}.
\end{equation}
Relating the chemical potential of each species to its activity coefficient, as was done for the binary alloy, gives the first element of the transformed thermodynamic factor:
\begin{equation}
  \label{eq:general_11_element_thermodynamic_factor}
  \vec{1}^{\,\mathrm{T}} \tilde{\mathbf{\Theta}}_{\mathbf{x}} \vec{1} = \frac{c^{2}}{x_{\mathrm{Va}}} + \sum_{i=1}^{c} \frac{1}{x_{i}} +
  \sum_{i,j} f_{ij} = \frac{c^{2}}{x_{\mathrm{Va}}} + \mathcal{O}(1)
\end{equation}

No remaining element of \cref{eq:general_transformed_thermodynamic_factor} contains $x_{\Va}$ explicitly, so none of them diverges as the vacancies become dilute. The vacancy nevertheless enters these elements in two ways. The first is the dependence of the elemental activity coefficients on the vacancy concentration, which is negligible because the vacancies are dilute. The second is the activity coefficient of the vacancy itself.

Following the steps that led to \cref{eq:theta_mix} as outlined in \cref{sec-app:multicomponent-alloy-projection} gives, for $i,j=2,\ldots,c$:
\begin{align}
  \label{eq:general_projected_thermodynamic_factor}
  k_\mathrm{B} T(\hat{\mathbf{Q}}^\mathrm{T} \tilde{\mathbf{\Theta}}_{\mathbf{x}} \hat{\mathbf{Q}})_{ij}
  &= \left(\frac{\partial \hat{\mu}_{i}}{\partial x_{j}}\right)_{x_{k\neq j}} - \left(\frac{\partial \hat{\mu}_{i}}{\partial x_{1}}\right)_{x_{k\neq 1}} \nonumber \\
  &= \left(\frac{\partial \hat{\mu}_{i}}{\partial x_{j}}\right)_{x_{k\neq j,1},x_{\Va}} \approx \left(\frac{\partial^{2} g}{\partial y_{i}\partial y_{j}}\right)_{x_{k\neq j,1},x_{\Va}}
\end{align}
where $\hat{\mu}_{i} = \mu_{i} - \mu_{1}$. In each derivative, the composition that is neither differentiated nor held fixed is the dependent variable, determined by the constraint $\sum_{i}x_{i}+x_{\Va}=1$. In the first expression, $x_{\Va}$ is the dependent variable and is therefore allowed to vary, recovering the definition in \cref{eq:thermodynamic_factor}. In the final expression, $x_{\Va}$ is held fixed, making $x_{1}$ the dependent variable. In the dilute-vacancy limit, \Cref{eq:general_projected_thermodynamic_factor} identifies the lower right block of \cref{eq:general_transformed_thermodynamic_factor} as the thermodynamic factor of the Gibbs free energy of the vacancy-free alloy, evaluated along the alloy composition axis. The alloy thermodynamic factor has elements $k_\mathrm{B}T\,\tilde{\Theta}_{\mathrm{alloy},ij} \approx \left(\frac{\partial^{2} g}{\partial y_{i}\partial y_{j}}\right)_{x_{k\neq j,1},x_{\Va}}$ for $i,j=2,\ldots,c$, the second derivative of $g$ with respect to the alloy composition coordinates $y_{i}$ and $y_{j}$, taken at fixed vacancy composition and fixed $x_{k\neq j,1}$, with $x_{1}$ varying as the dependent variable. It therefore contains only the curvature of the vacancy-free alloy. Within this limit, the off-diagonal blocks can be discarded as well, since \cref{sec-app:depend-eigensp-diff} shows that setting them to zero leaves the eigenspectrum of the diffusion matrix unchanged. Together, these results give the approximation:
\begin{equation}
  \label{eq:general_approximate_transformed_thermodynamic_factor}
  \tilde{\mathbf{\Theta}}_{\mathbf{y}} \approx
  \begin{bmatrix}
    \frac{c^{2}}{x_{\Va}}      &
    \vec{0}^{\mathrm{T}}          \\
    \vec{0} &
    \mathbf{\tilde{\Theta}}_{\mathrm{alloy}}
  \end{bmatrix}
\end{equation}
where $\vec{0}$ is a vector of dimension $(c-1)$ containing zeros.

Both elements of \cref{eq:general_approximate_transformed_thermodynamic_factor} can be computed through semi-grand canonical Monte Carlo simulations or through simpler approximations. The vacancy concentration follows from the scheme proposed by \citeauthor{lee_modeling_2026} \cite{lee_modeling_2026} and \citeauthor{belak_effect_2015} \cite{belak_effect_2015}, which applies to either canonical or semi-grand canonical Monte Carlo simulations. The alloy thermodynamic factor is most easily obtained from a semi-grand canonical simulation of the vacancy-free alloy, in which the exchange chemical potentials $\hat{\mu}_{i}$ are set so that the average composition of the alloy is $x_{i}$ and vacancies are never allowed to enter or leave the system. The composition fluctuations sampled in that simulation give the inverse of the alloy thermodynamic factor directly \cite{van_der_ven_vacancy_2010}:
\begin{equation}
  \label{eq:vacancy_free_covariance}
  \tilde{\Theta}^{-1}_{\mathrm{alloy},mn} = \frac{1}{M} \left( \langle N_{m}N_{n} \rangle -
  \langle N_{m} \rangle \langle N_{n} \rangle \right).
\end{equation}

\noindent Here $N_{m}$ is the number of atoms of species $m$ distributed over the $M$ substitutional sites, and the averages are taken over the sampled configurations. $\mathbf{\tilde{\Theta}}_{\mathrm{alloy}}$ is obtained by inverting the matrix constructed from \cref{eq:vacancy_free_covariance} with entries corresponding to the dependent species $x_1$ removed. Subsequently, the result can be inserted into \cref{eq:general_approximate_transformed_thermodynamic_factor} to obtain $\tilde{\mathbf{\Theta}}_{\mathbf{y}}$. A coordinate transform back to the original basis yields the approximate thermodynamic factor $\tilde{\mathbf{\Theta}}_{\mathbf{x}}$.

\subsection{Thermodynamic factor from a single-vacancy expansion}
\label{sec:cg_derivation}
The formalism of the previous sections approximates the thermodynamic factor in the dilute-vacancy limit. It captures the relevant diffusion physics but is not numerically exact. Here we present a second route that recovers the full thermodynamic factor of the vacancy-containing alloy, without setting any of its elements to zero. This route is exact to within a truncation of the semi-grand canonical partition function at a single vacancy, and serves as an independent benchmark for the approximation developed above. \Cref{sec-app:cg_derivation} gives the full derivation, and only the resulting expressions are described here.

\Cref{eq:vacancy_free_covariance} requires only the $(c-1)\times(c-1)$ block of covariances among the elements other than the dependent species, species 1. The thermodynamic factor of the vacancy-containing alloy is instead a $c \times c$ matrix, and recovering it in full requires the covariances of all $c$ elements. That $c \times c$ covariance matrix is singular for a vacancy-free alloy and therefore cannot be inverted. To circumvent this singularity, we extend a previously developed framework for sampling dilute vacancies~\cite{belak_effect_2015,lee_modeling_2026} into a single-vacancy expansion of the covariance matrix, readjusting the composition fluctuations to account for vacancies as:

\begin{align}
    \langle N_i N_j \rangle 
	&= \frac{\langle N_{i}(\vec{\sigma})N_{j}(\vec{\sigma}) + 
    \frac{1}{c}\sum_{\vec{\nu}} 
    N_{i}(\vec{\nu})N_{j}(\vec{\nu})
    \exp(-\beta \Delta \Omega(\vec{\nu}))
    \rangle}
    {1+\xi}
    \label{eq:product_term_general} \\
    \langle N_i \rangle &= \frac{\langle N_{i}(\vec{\sigma}) + 
    \frac{1}{c}\sum_{\vec{\nu}} 
    N_{i}(\vec{\nu})
    \exp(-\beta \Delta \Omega(\vec{\nu})) \rangle} 
    {1+\xi}
    \label{eq:mean_term_general}
\end{align}
Here $\vec{\sigma}$ denotes a vacancy-free configuration sampled during the semi-grand canonical Monte Carlo (GCMC) simulation, and $\vec{\nu}$ a configuration obtained by replacing a single atom of $\vec{\sigma}$ with a vacancy. For each sampled $\vec{\sigma}$, every lattice site is considered in turn as a candidate vacancy site, and the sums in \cref{eq:product_term_general,eq:mean_term_general} run over the resulting set of configurations. Each configuration is weighted by $\exp(-\beta \Delta \Omega(\vec{\nu}))$, where $\Delta\Omega(\vec{\nu})$ is the change in semi-grand potential on introducing the vacancy. The quantity $\xi$, given in \cref{eq:xi}, is the ensemble average of the vacancy partition function of \cref{eq:vacancy_partition_function}. The resulting single-vacancy covariance matrix is non-singular and can therefore be inverted to obtain the thermodynamic factor directly.

\section{Results}
\subsection{Benchmarks of the thermodynamic factor computed from the single-vacancy expansion}
\label{sec:benchm-therm-coarse}

\begin{figure}
	\centering
	\includegraphics{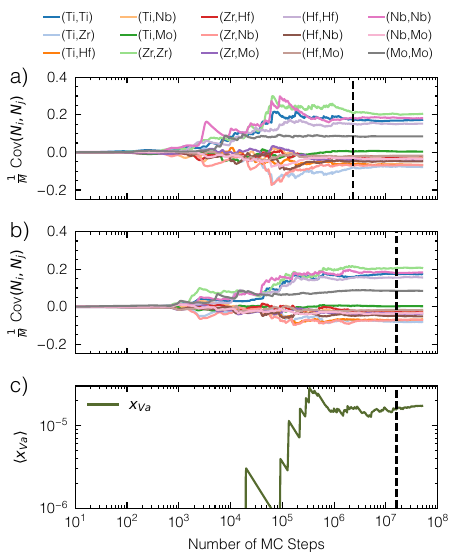}
	\caption{Convergence of covariance values as a function of the number of GCMC steps for (a) vacancy-free HfMoNbTiZr and (b) HfMoNbTiZr with vacancies. (c) Convergence of the vacancy concentration in HfMoNbTiZr. All simulations are at 2000~K. In panels (a) and (b), each line corresponds to the covariance between a distinct pair of elements, with the corresponding legend at the top of the figure.}
	\label{fig:convergence}
\end{figure}

We first demonstrate the accuracy of the single-vacancy expansion using the prototypical Senkov alloy HfMoNbTiZr~\cite{senkov_mechanical_2011}. This quinary alloy forms a single-phase disordered solid solution on the bcc crystal at elevated temperatures~\cite{senkov_mechanical_2011,lee_modeling_2026}. We employed an eCE model~\cite{muller_constructing_2025,lee_modeling_2026} trained and benchmarked on vacancy energetics in multicomponent alloys of elements in groups 4--6 of the periodic table. Details of data acquisition, DFT calculations, and model parametrization can be found in \cite{lee_modeling_2026}. At 2000~K, the Senkov alloy exhibits a vacancy concentration of $1.7 \times 10^{-5}$, which is sufficiently high to allow reliable sampling with a moderate number of Monte Carlo steps. To compare the thermodynamic factors of the vacancy-free and vacancy-containing systems, we computed covariance matrices of composition fluctuations (\cref{eq:vacancy_free_covariance}) using semi-grand Canonical simulations (GCMC). \Cref{fig:convergence}a and \cref{fig:convergence}b show the convergence of the covariance elements against the number of GCMC steps. The covariances of the vacancy-free system converged after approximately $2\times10^6$ steps, whereas the vacancy-containing system required roughly an order of magnitude more. This slow convergence arises from the low vacancy concentration, which demands more Monte Carlo steps to achieve adequate sampling. Once the vacancy concentration itself converges (\cref{fig:convergence}c), the covariance values also stabilize. The vacancy concentration of this alloy is high compared with other refractory alloys, where it can be as low as $10^{-10}$ even at elevated temperatures~\cite{lee_modeling_2026}. Lower vacancy concentrations will require even longer simulations to resolve the elements of the covariance matrix with sufficient numerical accuracy.

\begin{figure}
	\centering
	\includegraphics{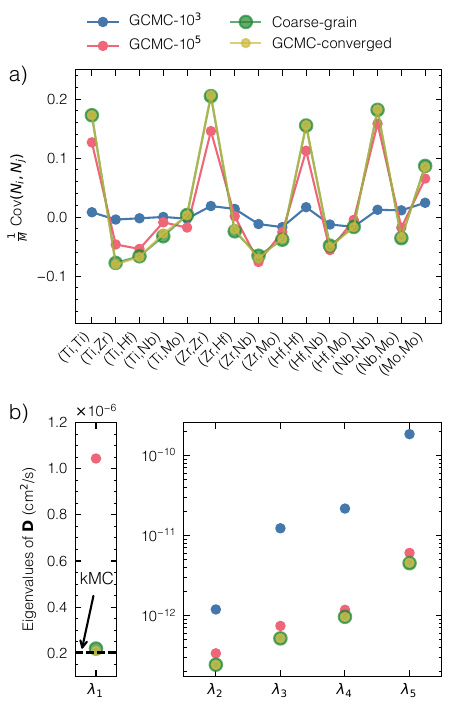}
	\caption{(a) Covariance values for each pair of elements in HfMoNbTiZr at 2000~K, comparing fully converged GCMC simulations (ground truth) against undersampled simulations using $10^3$ and $10^5$ MC samples and against the single-vacancy expansion. (b) Eigenvalues of $\mathbf{D}$ obtained from each of the covariance matrices in (a).}
	\label{fig:covariance_comparison}
\end{figure}

\Cref{fig:covariance_comparison} compares the covariance matrix elements evaluated through several simulation strategies. We computed the elements of $\mathbf{\tilde{\Theta}}^{-1}$ using \cref{eq:vacancy_free_covariance} from fully converged GCMC simulations in which vacancies enter and leave the system, and from the single-vacancy expansion of \cref{eq:mean_term_general,eq:product_term_general}. For comparison, we also evaluated covariance matrices from $10^3$ and $10^5$ samples drawn from the same vacancy-containing GCMC trajectory, denoted GCMC-$10^3$ and GCMC-$10^5$, respectively. \Cref{fig:covariance_comparison}a presents the upper triangular elements of $\mathbf{\tilde{\Theta}}^{-1}$. Covariances from GCMC-$10^3$ deviated significantly from the converged values. The GCMC-$10^5$ results showed improved agreement, but noticeable discrepancies remained. In contrast, the single-vacancy covariances closely matched the converged values. The single-vacancy expansion thus achieves reliable estimates at roughly one-tenth the computational cost.

\subsection{Benchmarking the effects of approximating the thermodynamic factor on the eigenspectrum of $\mathbf{D}$}
\label{sec:benchm-effects-appr}

The covariance matrix of \cref{eq:vacancy_free_covariance} and its inverse, the thermodynamic factor, are used to compute the matrix of non-dilute diffusion coefficients in concentrated alloys. To quantify the impact of errors in $\mathbf{\tilde{\Theta}}^{-1}$ on diffusion properties, we computed $\mathbf{D}$ using the covariance matrices from \cref{sec:cg_derivation}. The Onsager transport coefficients $\mathbf{\tilde{L}}$ for the HfMoNbTiZr alloy at 2000~K were obtained from kinetic Monte Carlo (kMC) simulations with migration barriers for nearest-neighbor vacancy hops fixed at their pure-element values~\cite{shang_comprehensive_2016}. The migration barriers for Nb and Mo were taken from their bcc phases, whereas those for Ti, Zr, and Hf were taken from their hcp phases. Although migration barriers in real multicomponent alloys depend on the local chemical environment~\cite{van_der_ven_first-principles_2001,van_der_ven_first_2005,goiri_role_2019,behara_role_2024,kehr_mobility_1989,mishin_monte_1997,lee_diffusion_2026}, we adopted this simplified model to isolate the effect of $\mathbf{\tilde{\Theta}}^{-1}$ on $\mathbf{D}$.

We analyzed the eigenspectrum of $\mathbf{D}$ computed with the thermodynamic factor estimated from GCMC and the single-vacancy expansion developed in this study. From \cref{eq:eigenvalue_binary_plus} the largest eigenvalue ($\lambda_1$) can be related to the vacancy tracer diffusion coefficient, while the remaining eigenvalues correspond to interdiffusion coefficients in the crystal frame of reference~\cite{kehr_mobility_1989,van_der_ven_vacancy_2010}. \Cref{fig:covariance_comparison}b compares the full eigenspectrum of $\mathbf{D}$ for the quinary Senkov alloy across the various methods used to estimate the thermodynamic factor, and the largest eigenvalue with the vacancy tracer diffusion coefficient ($D^*_{\mathrm{Va}}$) obtained from kMC.

\Cref{fig:covariance_comparison}b shows that the fully converged GCMC simulations and the single-vacancy expansion yield the largest eigenvalue, $\lambda_1$, in close agreement with $D^*_{\mathrm{Va}}$. Using $10^5$ samples overestimated $\lambda_1$ by one order of magnitude, because the shorter Monte Carlo trajectories estimate the vacancy concentration poorly. Using $10^3$ samples overestimated it by approximately six orders of magnitude, placing the corresponding $\lambda_1$ outside the range of the plot. This demonstrates how small errors in the covariance matrix can be strongly amplified upon inversion and propagate into $\mathbf{D}$. The smaller eigenvalues ($\lambda_{2}$--$\lambda_{5}$) are less sensitive to numerical noise. The single-vacancy values agree closely with those from fully converged GCMC simulations. In contrast, both GCMC-$10^5$ and GCMC-$10^3$ deviate from the converged results, with discrepancies of a factor of 2--3 for GCMC-$10^5$ and up to two orders of magnitude for GCMC-$10^3$.

\begin{figure}
	\centering
	\includegraphics{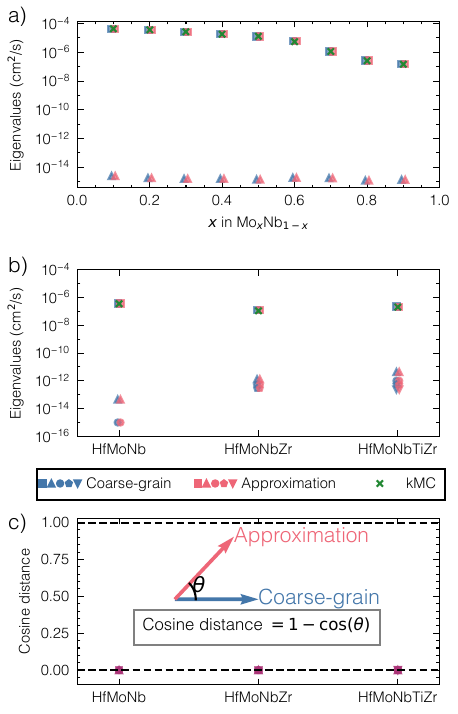}
	\caption{(a) Eigenvalues of $\mathbf{D}$ in Mo$_x$Nb$_{1-x}$ as a function of composition. (b) Eigenvalues of $\mathbf{D}$ in HfMoNb, HfMoNbZr and HfMoNbTiZr, obtained using the $\mathbf{\tilde{\Theta}}$ matrices computed from the single-vacancy expansion and from the approximation of \cref{eq:general_approximate_transformed_thermodynamic_factor}. The value of $D^*_\mathrm{Va}$ obtained from kMC is shown as green crosses. Values on the $x$-axis are shifted slightly to separate the two methods. (c) Cosine distances between the eigenvectors computed from the single-vacancy expansion and the approximation of \cref{eq:general_approximate_transformed_thermodynamic_factor} for the three alloys in (b).}
	\label{fig:eig_comparison}
\end{figure}

The single-vacancy expansion reproduces the thermodynamic factor of fully converged GCMC simulations at a fraction of the cost. It therefore serves as a benchmark for the approximate thermodynamic factor of \cref{eq:general_approximate_transformed_thermodynamic_factor}, in which the curvature of the free energy of the vacancy-free alloy and the vacancy concentration are the only inputs. We computed the eigenvalues of $\mathbf{D}$ from both routes for a range of alloys within the HfMoNbTiZr composition space. \Cref{fig:eig_comparison}a compares the eigenvalues for Mo$_x$Nb$_{1-x}$ as a function of composition, and \cref{fig:eig_comparison}b the eigenspectra of HfMoNb, HfMoNbZr, and HfMoNbTiZr. All eigenvalues are reproduced by both the single-vacancy thermodynamic factor and the approximation of \cref{eq:general_approximate_transformed_thermodynamic_factor}, and the largest eigenvalue agrees with the vacancy tracer diffusion coefficient computed directly from kMC. The approximation therefore reproduces the dominant diffusion behavior while avoiding both the single-vacancy expansion and long-running GCMC simulations, providing a simple and computationally efficient route for estimating the eigenspectrum of $\mathbf{D}$.

The agreement extends beyond the eigenvalues to the corresponding eigenvectors of $\mathbf{D}$. While the eigenvalues determine the characteristic diffusion rates, the eigenvectors define the independent diffusion modes, describing the coupled motion of the constituent elements. We quantified the similarity between the eigenvectors from the two routes using the cosine distance $1-\cos(\theta)$, where $\theta$ is the angle between the corresponding eigenvectors. \Cref{fig:eig_comparison}c displays the cosine distances for the HfMoNb, HfMoNbZr, and HfMoNbTiZr alloys. We found cosine distances near zero across all systems considered, indicating that the eigenvectors from the two methods are aligned. The residual errors are of the order of the vacancy concentration, as expected from the derivation in \cref{sec-app:depend-eigensp-diff}. The approximation of \cref{eq:general_approximate_transformed_thermodynamic_factor} therefore preserves not only the magnitude of the interdiffusion coefficients but also the dominant directions of coupled diffusion in composition space.

\section{Discussion}
\label{sec:discussion}
Converting Onsager transport coefficients into diffusion coefficients requires the thermodynamic factor, and in a substitutional alloy, that factor must be evaluated for an alloy carrying only a trace of vacancies. The resulting matrix is nearly singular, and the composition fluctuations that determine it converge slowly in Monte Carlo simulations. Rotating the composition axes so that one axis measures the vacancy concentration and the remaining $c-1$ axes lie in the vacancy-free hyperplane separates the thermodynamic factor into contributions that can be evaluated independently. In the dilute vacancy limit, one diagonal element, $c^{2}/x_{\Va}$, carries the entire divergence and depends on nothing but the vacancy concentration. The remaining block is the curvature of the free energy of the vacancy-free alloy, $\mathbf{\tilde{\Theta}}_{\mathrm{alloy}}$, which an ordinary simulation without vacancies supplies. The off-diagonal blocks, which hold the interaction between the vacancy and the alloy, do not reach the eigenvalues of $\mathbf{D}$ until second order in the vacancy concentration and can be discarded. What remains, \cref{eq:general_approximate_transformed_thermodynamic_factor}, is a thermodynamic factor assembled from two ingredients that are each straightforward to obtain: the vacancy concentration, and the free-energy curvature of the vacancy-free alloy.

We have developed two practical methods for computing the thermodynamic factor in the dilute-vacancy limit. Direct evaluation of $\mathbf{\tilde{\Theta}}$ is numerically challenging because it is ill-conditioned, making accurate computation from composition fluctuations difficult. Both methods circumvent this issue while requiring only composition fluctuations obtained from GCMC simulations without vacancies. The first exploits a linear transformation of the composition space to separate the vacancy contribution from the elemental degrees of freedom, reconstructing an approximation of $\mathbf{\tilde{\Theta}}$ (\cref{eq:general_approximate_transformed_thermodynamic_factor}) using only the vacancy-free alloy and the equilibrium vacancy concentration. When numerically accurate values are required, the second method extends a previously developed sampling framework~\cite{belak_effect_2015,lee_modeling_2026} into a single-vacancy expansion that explicitly accounts for vacancies in the covariance matrix. We demonstrated that both approaches closely reproduce the thermodynamic factor obtained from full GCMC, while substantially reducing the computational cost. Moreover, the resulting eigenspectra of $\mathbf{D}$ are in agreement with direct calculations, demonstrating that these approximations preserve the physically relevant thermodynamic contributions to the diffusion coefficients and provide an efficient route for high-fidelity diffusion simulations in compositionally complex alloys.

The same separation settles the structure of the eigenspectrum of $\mathbf{D}$. \citeauthor{van_der_ven_vacancy_2010} showed for a binary alloy that the larger of the two eigenvalues is independent of both the vacancy concentration and the thermodynamic factor~\cite{van_der_ven_vacancy_2010}. \Cref{sec-app:depend-eigensp-diff} extends this to an alloy of $c$ components with arbitrary thermodynamics. Writing $\mathbf{\tilde{L}}_{\mathbf{y}} = x_{\Va}\boldsymbol{\Lambda}$ and separating the divergent element of $\mathbf{\tilde{\Theta}}_{\mathbf{y}}$ leaves a characteristic equation whose largest root is $\lambda_{1} = D^{\star}_{\Va} + \mathcal{O}(x_{\Va})$. To our knowledge this is the first demonstration that the largest eigenvalue of the diffusion matrix is the vacancy tracer diffusion coefficient for any number of components and any degree of non-ideality. The remaining $c-1$ eigenvalues scale linearly with the vacancy concentration and are set entirely by $\mathbf{\tilde{\Theta}}_{\mathrm{alloy}}$. \Cref{fig:eig_comparison} numerically verifies this, with $\lambda_{1}$ coinciding with $D^{\star}_{\Va}$ obtained independently from kMC across every alloy considered.

\Cref{eq:general_approximate_transformed_thermodynamic_factor} also lifts the requirement that the thermodynamic factor come from a first-principles model. Its second ingredient, $\mathbf{\tilde{\Theta}}_{\mathrm{alloy}}$, is the curvature of the free energy of the vacancy-free alloy, which is what a CALPHAD assessment already provides. Wherever an assessed description is valid across the composition domain of interest, $\mathbf{\tilde{\Theta}}_{\mathrm{alloy}}$ follows from the Hessian of the assessed free energy, leaving the vacancy concentration as the only remaining quantity. That concentration can be obtained from the techniques of \citeauthor{belak_effect_2015} and \citeauthor{lee_modeling_2026}~\cite{belak_effect_2015,lee_modeling_2026}, which require only a simulation of the vacancy-free alloy. For a first pass, a rule-of-mixtures estimate built from the vacancy formation energies of the pure elements may place the diffusion coefficients in the right range. Such an estimate should be treated as a ballpark figure. It will not capture how the vacancy concentration varies with alloy composition, and it should not be used to extract chemical trends.

\begin{figure}
	\centering
	\includegraphics{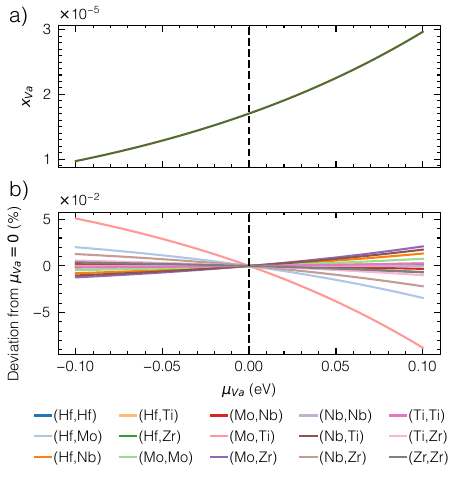}
	\caption{(a) Vacancy concentration and (b) covariance values for each distinct pair of elements in HfMoNbTiZr at 2000~K, as the vacancy chemical potential is displaced from $\mu_{\mathrm{Va}} = 0$. Panel (a) shows the vacancy concentration itself, panel (b) the difference relative to $\mu_{\mathrm{Va}} = 0$.}
	\label{fig:mu_scaling}
\end{figure}

Our analysis thus far has been restricted to a system where the vacancy chemical potential is zero. In real materials, vacancy sources and sinks such as grain boundaries and dislocations are typically assumed to be present throughout the material to regulate the vacancy concentration at its equilibrium value~\cite{van_der_ven_vacancy_2010}. However, at the continuum scale, non-equilibrium situations may arise for several reasons, such as when a vacancy migrates far from such sources and sinks, causing a local deviation of $\mu_\mathrm{Va}$ from zero~\cite{yu_theory_2008,soisson_cu-precipitation_2007,nastar_atomistic_2012}. In such scenarios, the diffusion matrix, the thermodynamic factor, and the Onsager transport coefficients must all be recomputed at the local vacancy concentration. The Onsager transport coefficients can, to first order, be rescaled by the vacancy concentration. The thermodynamic factor requires more care. As derived in \cref{sec-app:cg_derivation} (\cref{eqn:mean_term_with_muVa,eqn:product_term_with_muVa}), the individual terms in the ensemble averages of \cref{eq:product_term_vacancy_equilibrium,eq:mean_term_vacancy_equilibrium} can be rescaled by $\exp(\beta \mu_\mathrm{Va})$, where $\mu_\mathrm{Va}$ is the local vacancy chemical potential corresponding to a vacancy concentration that deviates from $x_\mathrm{Va}(\mu_\mathrm{Va}=0)$.

\Cref{fig:mu_scaling} shows the resulting vacancy concentration and covariance values for $\mu_\mathrm{Va}$ deviating from zero for the same HfMoNbTiZr alloy at 2000~K. The vacancy concentration changes by approximately a factor of two when $\mu_\mathrm{Va}$ is shifted by 0.1~eV. In contrast, the covariance matrix elements display a much weaker response, with relative changes on the order of 0.1\% compared to their equilibrium values. This indicates that local deviations in vacancy chemical potential primarily affect the vacancy concentration. Since the thermodynamic factor changes only marginally, the dominant effect on the smaller eigenvalues is due to the variation in vacancy concentration. \Cref{eq:product_term_vacancy_equilibrium,eq:mean_term_vacancy_equilibrium} provide a scaling approach to correct for the vacancy concentration and covariance matrices when vacancies are in non-equilibrium. Alternatively, $\mathbf{\tilde{\Theta}}$ can also be approximated from \cref{eq:general_approximate_transformed_thermodynamic_factor} using only the change in vacancy concentration in the first diagonal entry. These proposed scaling approaches provide a computationally efficient means to capture these effects without requiring additional simulations. This is particularly useful in continuum simulations where local regions deviate from the continuous vacancy-source/sink approximation, as both $\mathbf{\tilde{L}}$ and $\mathbf{\tilde{\Theta}}$ can be rescaled consistently to account for the local vacancy concentration.

\section{Conclusion}
We developed two routes to the thermodynamic factor of a multicomponent alloy in the dilute-vacancy limit, both requiring only simulations of the vacancy-free alloy. A transformation of the composition axes confines the vacancy contribution to a single divergent term and leaves a remainder set by the free-energy curvature of the vacancy-free alloy, giving the approximation of \cref{eq:general_approximate_transformed_thermodynamic_factor}. Truncating the semi-grand canonical partition function at a single vacancy recovers the full thermodynamic factor and provides an independent benchmark. Both reproduce the thermodynamic factor of fully converged GCMC simulations at roughly one tenth of the sampling effort. The same separation shows that the largest eigenvalue of the diffusion matrix is the vacancy tracer diffusion coefficient for any number of components and any degree of non-ideality, while the remaining $c-1$ eigenvalues scale with the vacancy concentration. Because \cref{eq:general_approximate_transformed_thermodynamic_factor} requires only the vacancy concentration and the free-energy curvature of the vacancy-free alloy, thermodynamic descriptions from atomistic models or CALPHAD assessments can be used directly in mesoscale simulations of mass transport. A rescaling relation extends these results to local vacancy chemical potentials away from equilibrium, without additional simulations.

\begin{acknowledgments}
  This research was supported by the Swiss National Science Foundation through grant number 215178. The authors also acknowledge access to Eiger at the Swiss National Supercomputing Centre with project ID mr30 under the NCCR MARVEL's share, a National Centre of Competence in Research funded by the Swiss National Science Foundation (grant number 205602).
\end{acknowledgments}

\bibliography{references}

\begin{thebibliography}{37}%
\makeatletter
\providecommand \@ifxundefined [1]{%
 \@ifx{#1\undefined}
}%
\providecommand \@ifnum [1]{%
 \ifnum #1\expandafter \@firstoftwo
 \else \expandafter \@secondoftwo
 \fi
}%
\providecommand \@ifx [1]{%
 \ifx #1\expandafter \@firstoftwo
 \else \expandafter \@secondoftwo
 \fi
}%
\providecommand \natexlab [1]{#1}%
\providecommand \enquote  [1]{``#1''}%
\providecommand \bibnamefont  [1]{#1}%
\providecommand \bibfnamefont [1]{#1}%
\providecommand \citenamefont [1]{#1}%
\providecommand \href@noop [0]{\@secondoftwo}%
\providecommand \href [0]{\begingroup \@sanitize@url \@href}%
\providecommand \@href[1]{\@@startlink{#1}\@@href}%
\providecommand \@@href[1]{\endgroup#1\@@endlink}%
\providecommand \@sanitize@url [0]{\catcode `\\12\catcode `\$12\catcode `\&12\catcode `\#12\catcode `\^12\catcode `\_12\catcode `\%12\relax}%
\providecommand \@@startlink[1]{}%
\providecommand \@@endlink[0]{}%
\providecommand \url  [0]{\begingroup\@sanitize@url \@url }%
\providecommand \@url [1]{\endgroup\@href {#1}{\urlprefix }}%
\providecommand \urlprefix  [0]{URL }%
\providecommand \Eprint [0]{\href }%
\providecommand \doibase [0]{https://doi.org/}%
\providecommand \selectlanguage [0]{\@gobble}%
\providecommand \bibinfo  [0]{\@secondoftwo}%
\providecommand \bibfield  [0]{\@secondoftwo}%
\providecommand \translation [1]{[#1]}%
\providecommand \BibitemOpen [0]{}%
\providecommand \bibitemStop [0]{}%
\providecommand \bibitemNoStop [0]{.\EOS\space}%
\providecommand \EOS [0]{\spacefactor3000\relax}%
\providecommand \BibitemShut  [1]{\csname bibitem#1\endcsname}%
\let\auto@bib@innerbib\@empty
\bibitem [{\citenamefont {Darken}(1951)}]{darken_formal_1951}%
  \BibitemOpen
  \bibfield  {author} {\bibinfo {author} {\bibfnamefont {L.~S.}\ \bibnamefont {Darken}},\ }\bibfield  {title} {\bibinfo {title} {Formal basis of diffusion theory},\ }\href@noop {} {\bibfield  {journal} {\bibinfo  {journal} {Atom movements}\ ,\ \bibinfo {pages} {1}} (\bibinfo {year} {1951})}\BibitemShut {NoStop}%
\bibitem [{\citenamefont {Allnatt}\ and\ \citenamefont {Lidiard}(2003)}]{allnatt_atomic_2003}%
  \BibitemOpen
  \bibfield  {author} {\bibinfo {author} {\bibfnamefont {A.~R.}\ \bibnamefont {Allnatt}}\ and\ \bibinfo {author} {\bibfnamefont {A.~B.}\ \bibnamefont {Lidiard}},\ }\href@noop {} {\emph {\bibinfo {title} {Atomic {Transport} in {Solids}}}}\ (\bibinfo  {publisher} {Cambridge University Press},\ \bibinfo {year} {2003})\BibitemShut {NoStop}%
\bibitem [{\citenamefont {Balluffi}\ \emph {et~al.}(2005)\citenamefont {Balluffi}, \citenamefont {Allen},\ and\ \citenamefont {Carter}}]{balluffi_kinetics_2005}%
  \BibitemOpen
  \bibfield  {author} {\bibinfo {author} {\bibfnamefont {R.~W.}\ \bibnamefont {Balluffi}}, \bibinfo {author} {\bibfnamefont {S.~M.}\ \bibnamefont {Allen}},\ and\ \bibinfo {author} {\bibnamefont {Carter}},\ }\href {https://onlinelibrary.wiley.com/doi/abs/10.1002/0471749311.ch1} {\emph {\bibinfo {title} {Kinetics of {Materials}}}}\ (\bibinfo  {publisher} {John Wiley \& Sons, Ltd},\ \bibinfo {year} {2005})\BibitemShut {NoStop}%
\bibitem [{\citenamefont {Mishin}\ and\ \citenamefont {Farkas}(1997{\natexlab{a}})}]{mishin_atomistic_1997}%
  \BibitemOpen
  \bibfield  {author} {\bibinfo {author} {\bibfnamefont {Y.}~\bibnamefont {Mishin}}\ and\ \bibinfo {author} {\bibfnamefont {D.}~\bibnamefont {Farkas}},\ }\bibfield  {title} {\bibinfo {title} {Atomistic simulation of point defects and diffusion in {B2} {NiAl}: {Part} {II}. {Diffusion} mechanisms},\ }\href {https://doi.org/10.1080/01418619708210290} {\bibfield  {journal} {\bibinfo  {journal} {Philosophical Magazine A}\ }\textbf {\bibinfo {volume} {75}},\ \bibinfo {pages} {187} (\bibinfo {year} {1997}{\natexlab{a}})}\BibitemShut {NoStop}%
\bibitem [{\citenamefont {Mishin}\ and\ \citenamefont {Farkas}(1997{\natexlab{b}})}]{mishin_atomistic_1997-1}%
  \BibitemOpen
  \bibfield  {author} {\bibinfo {author} {\bibfnamefont {Y.}~\bibnamefont {Mishin}}\ and\ \bibinfo {author} {\bibfnamefont {D.}~\bibnamefont {Farkas}},\ }\bibfield  {title} {\bibinfo {title} {Atomistic simulation of point defects and diffusion in {B2} {NiAl}: {Part} {I}. {Point} defect energetics},\ }\href {https://doi.org/10.1080/01418619708210289} {\bibfield  {journal} {\bibinfo  {journal} {Philosophical Magazine A}\ }\textbf {\bibinfo {volume} {75}},\ \bibinfo {pages} {169} (\bibinfo {year} {1997}{\natexlab{b}})}\BibitemShut {NoStop}%
\bibitem [{\citenamefont {Kehr}\ \emph {et~al.}(1989)\citenamefont {Kehr}, \citenamefont {Binder},\ and\ \citenamefont {Reulein}}]{kehr_mobility_1989}%
  \BibitemOpen
  \bibfield  {author} {\bibinfo {author} {\bibfnamefont {K.~W.}\ \bibnamefont {Kehr}}, \bibinfo {author} {\bibfnamefont {K.}~\bibnamefont {Binder}},\ and\ \bibinfo {author} {\bibfnamefont {S.~M.}\ \bibnamefont {Reulein}},\ }\bibfield  {title} {\bibinfo {title} {Mobility, interdiffusion, and tracer diffusion in lattice-gas models of two-component alloys},\ }\href {https://doi.org/10.1103/PhysRevB.39.4891} {\bibfield  {journal} {\bibinfo  {journal} {Physical Review B}\ }\textbf {\bibinfo {volume} {39}},\ \bibinfo {pages} {4891} (\bibinfo {year} {1989})}\BibitemShut {NoStop}%
\bibitem [{\citenamefont {Van~der Ven}\ \emph {et~al.}(2010)\citenamefont {Van~der Ven}, \citenamefont {Yu}, \citenamefont {Ceder},\ and\ \citenamefont {Thornton}}]{van_der_ven_vacancy_2010}%
  \BibitemOpen
  \bibfield  {author} {\bibinfo {author} {\bibfnamefont {A.}~\bibnamefont {Van~der Ven}}, \bibinfo {author} {\bibfnamefont {H.-C.}\ \bibnamefont {Yu}}, \bibinfo {author} {\bibfnamefont {G.}~\bibnamefont {Ceder}},\ and\ \bibinfo {author} {\bibfnamefont {K.}~\bibnamefont {Thornton}},\ }\bibfield  {title} {\bibinfo {title} {Vacancy mediated substitutional diffusion in binary crystalline solids},\ }\href {https://doi.org/10.1016/j.pmatsci.2009.08.001} {\bibfield  {journal} {\bibinfo  {journal} {Progress in Materials Science}\ }\textbf {\bibinfo {volume} {55}},\ \bibinfo {pages} {61} (\bibinfo {year} {2010})}\BibitemShut {NoStop}%
\bibitem [{\citenamefont {Cahn}\ and\ \citenamefont {Larché}(1983)}]{cahn_invariant_1983}%
  \BibitemOpen
  \bibfield  {author} {\bibinfo {author} {\bibfnamefont {J.~W.}\ \bibnamefont {Cahn}}\ and\ \bibinfo {author} {\bibfnamefont {F.~C.}\ \bibnamefont {Larché}},\ }\bibfield  {title} {\bibinfo {title} {An invariant formulation of multicomponent diffusion in crystals},\ }\href {https://doi.org/10.1016/0036-9748(83)90264-8} {\bibfield  {journal} {\bibinfo  {journal} {Scripta Metallurgica}\ }\textbf {\bibinfo {volume} {17}},\ \bibinfo {pages} {927} (\bibinfo {year} {1983})}\BibitemShut {NoStop}%
\bibitem [{\citenamefont {Yeh}\ \emph {et~al.}(2004)\citenamefont {Yeh}, \citenamefont {Chen}, \citenamefont {Lin}, \citenamefont {Gan}, \citenamefont {Chin}, \citenamefont {Shun}, \citenamefont {Tsau},\ and\ \citenamefont {Chang}}]{yeh_nanostructured_2004}%
  \BibitemOpen
  \bibfield  {author} {\bibinfo {author} {\bibfnamefont {J.-W.}\ \bibnamefont {Yeh}}, \bibinfo {author} {\bibfnamefont {S.-K.}\ \bibnamefont {Chen}}, \bibinfo {author} {\bibfnamefont {S.-J.}\ \bibnamefont {Lin}}, \bibinfo {author} {\bibfnamefont {J.-Y.}\ \bibnamefont {Gan}}, \bibinfo {author} {\bibfnamefont {T.-S.}\ \bibnamefont {Chin}}, \bibinfo {author} {\bibfnamefont {T.-T.}\ \bibnamefont {Shun}}, \bibinfo {author} {\bibfnamefont {C.-H.}\ \bibnamefont {Tsau}},\ and\ \bibinfo {author} {\bibfnamefont {S.-Y.}\ \bibnamefont {Chang}},\ }\bibfield  {title} {\bibinfo {title} {Nanostructured {High}-{Entropy} {Alloys} with {Multiple} {Principal} {Elements}: {Novel} {Alloy} {Design} {Concepts} and {Outcomes}},\ }\href {https://doi.org/10.1002/adem.200300567} {\bibfield  {journal} {\bibinfo  {journal} {Advanced Engineering Materials}\ }\textbf {\bibinfo {volume} {6}},\ \bibinfo {pages} {299} (\bibinfo {year} {2004})}\BibitemShut {NoStop}%
\bibitem [{\citenamefont {George}\ \emph {et~al.}(2019)\citenamefont {George}, \citenamefont {Raabe},\ and\ \citenamefont {Ritchie}}]{george_high-entropy_2019}%
  \BibitemOpen
  \bibfield  {author} {\bibinfo {author} {\bibfnamefont {E.~P.}\ \bibnamefont {George}}, \bibinfo {author} {\bibfnamefont {D.}~\bibnamefont {Raabe}},\ and\ \bibinfo {author} {\bibfnamefont {R.~O.}\ \bibnamefont {Ritchie}},\ }\bibfield  {title} {\bibinfo {title} {High-entropy alloys},\ }\href {https://doi.org/10.1038/s41578-019-0121-4} {\bibfield  {journal} {\bibinfo  {journal} {Nature Reviews Materials}\ }\textbf {\bibinfo {volume} {4}},\ \bibinfo {pages} {515} (\bibinfo {year} {2019})}\BibitemShut {NoStop}%
\bibitem [{\citenamefont {Miracle}\ and\ \citenamefont {Senkov}(2017)}]{miracle_critical_2017}%
  \BibitemOpen
  \bibfield  {author} {\bibinfo {author} {\bibfnamefont {D.~B.}\ \bibnamefont {Miracle}}\ and\ \bibinfo {author} {\bibfnamefont {O.~N.}\ \bibnamefont {Senkov}},\ }\bibfield  {title} {\bibinfo {title} {A critical review of high entropy alloys and related concepts},\ }\href {https://doi.org/10.1016/j.actamat.2016.08.081} {\bibfield  {journal} {\bibinfo  {journal} {Acta Materialia}\ }\textbf {\bibinfo {volume} {122}},\ \bibinfo {pages} {448} (\bibinfo {year} {2017})}\BibitemShut {NoStop}%
\bibitem [{\citenamefont {Tsai}\ \emph {et~al.}(2013)\citenamefont {Tsai}, \citenamefont {Tsai},\ and\ \citenamefont {Yeh}}]{tsai_sluggish_2013}%
  \BibitemOpen
  \bibfield  {author} {\bibinfo {author} {\bibfnamefont {K.~Y.}\ \bibnamefont {Tsai}}, \bibinfo {author} {\bibfnamefont {M.~H.}\ \bibnamefont {Tsai}},\ and\ \bibinfo {author} {\bibfnamefont {J.~W.}\ \bibnamefont {Yeh}},\ }\bibfield  {title} {\bibinfo {title} {Sluggish diffusion in {Co}–{Cr}–{Fe}–{Mn}–{Ni} high-entropy alloys},\ }\href {https://doi.org/10.1016/j.actamat.2013.04.058} {\bibfield  {journal} {\bibinfo  {journal} {Acta Materialia}\ }\textbf {\bibinfo {volume} {61}},\ \bibinfo {pages} {4887} (\bibinfo {year} {2013})}\BibitemShut {NoStop}%
\bibitem [{\citenamefont {Lee}\ and\ \citenamefont {Natarajan}(2026)}]{lee_diffusion_2026}%
  \BibitemOpen
  \bibfield  {author} {\bibinfo {author} {\bibfnamefont {D.~K.~J.}\ \bibnamefont {Lee}}\ and\ \bibinfo {author} {\bibfnamefont {A.~R.}\ \bibnamefont {Natarajan}},\ }\href {https://doi.org/10.48550/arXiv.2603.24228} {\bibinfo {title} {Diffusion coefficients of multi-principal element alloys from first principles}} (\bibinfo {year} {2026}),\ \bibinfo {note} {arXiv:2603.24228 [cond-mat]}\BibitemShut {NoStop}%
\bibitem [{\citenamefont {Daw}\ and\ \citenamefont {Chandross}(2021)}]{daw_sluggish_2021}%
  \BibitemOpen
  \bibfield  {author} {\bibinfo {author} {\bibfnamefont {M.~S.}\ \bibnamefont {Daw}}\ and\ \bibinfo {author} {\bibfnamefont {M.}~\bibnamefont {Chandross}},\ }\bibfield  {title} {\bibinfo {title} {Sluggish diffusion in random equimolar {FCC} alloys},\ }\href {https://doi.org/10.1103/PhysRevMaterials.5.043603} {\bibfield  {journal} {\bibinfo  {journal} {Physical Review Materials}\ }\textbf {\bibinfo {volume} {5}},\ \bibinfo {pages} {043603} (\bibinfo {year} {2021})}\BibitemShut {NoStop}%
\bibitem [{\citenamefont {Sen}\ \emph {et~al.}(2023)\citenamefont {Sen}, \citenamefont {Zhang}, \citenamefont {Rogal}, \citenamefont {Wilde}, \citenamefont {Grabowski},\ and\ \citenamefont {Divinski}}]{sen_anti-sluggish_2023}%
  \BibitemOpen
  \bibfield  {author} {\bibinfo {author} {\bibfnamefont {S.}~\bibnamefont {Sen}}, \bibinfo {author} {\bibfnamefont {X.}~\bibnamefont {Zhang}}, \bibinfo {author} {\bibfnamefont {L.}~\bibnamefont {Rogal}}, \bibinfo {author} {\bibfnamefont {G.}~\bibnamefont {Wilde}}, \bibinfo {author} {\bibfnamefont {B.}~\bibnamefont {Grabowski}},\ and\ \bibinfo {author} {\bibfnamefont {S.~V.}\ \bibnamefont {Divinski}},\ }\bibfield  {title} {\bibinfo {title} {‘{Anti}-sluggish’ {Ti} diffusion in {HCP} high-entropy alloys: {Chemical} complexity vs. lattice distortions},\ }\href {https://doi.org/10.1016/j.scriptamat.2022.115117} {\bibfield  {journal} {\bibinfo  {journal} {Scripta Materialia}\ }\textbf {\bibinfo {volume} {224}},\ \bibinfo {pages} {115117} (\bibinfo {year} {2023})}\BibitemShut {NoStop}%
\bibitem [{\citenamefont {Gao}\ \emph {et~al.}(2021)\citenamefont {Gao}, \citenamefont {Arróyave}, \citenamefont {Morral},\ and\ \citenamefont {Kattner}}]{gao_thermodynamics_2021}%
  \BibitemOpen
  \bibfield  {author} {\bibinfo {author} {\bibfnamefont {M.~C.}\ \bibnamefont {Gao}}, \bibinfo {author} {\bibfnamefont {R.}~\bibnamefont {Arróyave}}, \bibinfo {author} {\bibfnamefont {J.~E.}\ \bibnamefont {Morral}},\ and\ \bibinfo {author} {\bibfnamefont {U.~R.}\ \bibnamefont {Kattner}},\ }\bibfield  {title} {\bibinfo {title} {Thermodynamics and {Kinetics} of {High}-{Entropy} {Alloys}},\ }\href {https://doi.org/10.1007/s11669-021-00923-8} {\bibfield  {journal} {\bibinfo  {journal} {Journal of Phase Equilibria and Diffusion}\ }\textbf {\bibinfo {volume} {42}},\ \bibinfo {pages} {549} (\bibinfo {year} {2021})}\BibitemShut {NoStop}%
\bibitem [{\citenamefont {Kattner}(2020)}]{kattner_need_2020}%
  \BibitemOpen
  \bibfield  {author} {\bibinfo {author} {\bibfnamefont {U.~R.}\ \bibnamefont {Kattner}},\ }\bibfield  {title} {\bibinfo {title} {The need for reliable data in computational thermodynamics},\ }\bibfield  {journal} {\bibinfo  {journal} {High Temperatures - High Pressures}\ }\textbf {\bibinfo {volume} {49}},\ \href {https://doi.org/10.32908/hthp.v49.853} {10.32908/hthp.v49.853} (\bibinfo {year} {2020})\BibitemShut {NoStop}%
\bibitem [{\citenamefont {Wu}\ \emph {et~al.}(2022)\citenamefont {Wu}, \citenamefont {Kattner}, \citenamefont {Campbell}, \citenamefont {Guyer}, \citenamefont {Voorhees}, \citenamefont {Warren},\ and\ \citenamefont {Heinonen}}]{wu_co-based_2022}%
  \BibitemOpen
  \bibfield  {author} {\bibinfo {author} {\bibfnamefont {W.}~\bibnamefont {Wu}}, \bibinfo {author} {\bibfnamefont {U.~R.}\ \bibnamefont {Kattner}}, \bibinfo {author} {\bibfnamefont {C.~E.}\ \bibnamefont {Campbell}}, \bibinfo {author} {\bibfnamefont {J.~E.}\ \bibnamefont {Guyer}}, \bibinfo {author} {\bibfnamefont {P.~W.}\ \bibnamefont {Voorhees}}, \bibinfo {author} {\bibfnamefont {J.~A.}\ \bibnamefont {Warren}},\ and\ \bibinfo {author} {\bibfnamefont {O.~G.}\ \bibnamefont {Heinonen}},\ }\bibfield  {title} {\bibinfo {title} {Co-{Based} superalloy morphology evolution: {A} phase field study based on experimental thermodynamic and kinetic data},\ }\href {https://doi.org/10.1016/j.actamat.2022.117978} {\bibfield  {journal} {\bibinfo  {journal} {Acta Materialia}\ }\textbf {\bibinfo {volume} {233}},\ \bibinfo {pages} {117978} (\bibinfo {year} {2022})}\BibitemShut {NoStop}%
\bibitem [{\citenamefont {Belak}\ and\ \citenamefont {Van~der Ven}(2015)}]{belak_effect_2015}%
  \BibitemOpen
  \bibfield  {author} {\bibinfo {author} {\bibfnamefont {A.~A.}\ \bibnamefont {Belak}}\ and\ \bibinfo {author} {\bibfnamefont {A.}~\bibnamefont {Van~der Ven}},\ }\bibfield  {title} {\bibinfo {title} {Effect of disorder on the dilute equilibrium vacancy concentrations of multicomponent crystalline solids},\ }\href {https://doi.org/10.1103/PhysRevB.91.224109} {\bibfield  {journal} {\bibinfo  {journal} {Physical Review B}\ }\textbf {\bibinfo {volume} {91}},\ \bibinfo {pages} {224109} (\bibinfo {year} {2015})}\BibitemShut {NoStop}%
\bibitem [{\citenamefont {Goiri}\ \emph {et~al.}(2019)\citenamefont {Goiri}, \citenamefont {Kolli},\ and\ \citenamefont {Van~der Ven}}]{goiri_role_2019}%
  \BibitemOpen
  \bibfield  {author} {\bibinfo {author} {\bibfnamefont {J.~G.}\ \bibnamefont {Goiri}}, \bibinfo {author} {\bibfnamefont {S.~K.}\ \bibnamefont {Kolli}},\ and\ \bibinfo {author} {\bibfnamefont {A.}~\bibnamefont {Van~der Ven}},\ }\bibfield  {title} {\bibinfo {title} {Role of short- and long-range ordering on diffusion in {Ni}-{Al} alloys},\ }\href {https://doi.org/10.1103/PhysRevMaterials.3.093402} {\bibfield  {journal} {\bibinfo  {journal} {Physical Review Materials}\ }\textbf {\bibinfo {volume} {3}},\ \bibinfo {pages} {093402} (\bibinfo {year} {2019})}\BibitemShut {NoStop}%
\bibitem [{\citenamefont {Lee}\ \emph {et~al.}(2026)\citenamefont {Lee}, \citenamefont {Müller},\ and\ \citenamefont {Natarajan}}]{lee_modeling_2026}%
  \BibitemOpen
  \bibfield  {author} {\bibinfo {author} {\bibfnamefont {D.~K.~J.}\ \bibnamefont {Lee}}, \bibinfo {author} {\bibfnamefont {Y.~L.}\ \bibnamefont {Müller}},\ and\ \bibinfo {author} {\bibfnamefont {A.~R.}\ \bibnamefont {Natarajan}},\ }\bibfield  {title} {\bibinfo {title} {Modeling the equilibrium vacancy concentration in multi-principal element alloys from first-principles},\ }\href {https://doi.org/10.1016/j.actamat.2025.121752} {\bibfield  {journal} {\bibinfo  {journal} {Acta Materialia}\ }\textbf {\bibinfo {volume} {304}},\ \bibinfo {pages} {121752} (\bibinfo {year} {2026})}\BibitemShut {NoStop}%
\bibitem [{\citenamefont {Sanchez}\ \emph {et~al.}(1984)\citenamefont {Sanchez}, \citenamefont {Ducastelle},\ and\ \citenamefont {Gratias}}]{sanchez_generalized_1984}%
  \BibitemOpen
  \bibfield  {author} {\bibinfo {author} {\bibfnamefont {J.~M.}\ \bibnamefont {Sanchez}}, \bibinfo {author} {\bibfnamefont {F.}~\bibnamefont {Ducastelle}},\ and\ \bibinfo {author} {\bibfnamefont {D.}~\bibnamefont {Gratias}},\ }\bibfield  {title} {\bibinfo {title} {Generalized cluster description of multicomponent systems},\ }\href {https://doi.org/10.1016/0378-4371(84)90096-7} {\bibfield  {journal} {\bibinfo  {journal} {Physica A: Statistical Mechanics and its Applications}\ }\textbf {\bibinfo {volume} {128}},\ \bibinfo {pages} {334} (\bibinfo {year} {1984})}\BibitemShut {NoStop}%
\bibitem [{\citenamefont {Van~der Ven}\ \emph {et~al.}(2001)\citenamefont {Van~der Ven}, \citenamefont {Ceder}, \citenamefont {Asta},\ and\ \citenamefont {Tepesch}}]{van_der_ven_first-principles_2001}%
  \BibitemOpen
  \bibfield  {author} {\bibinfo {author} {\bibfnamefont {A.}~\bibnamefont {Van~der Ven}}, \bibinfo {author} {\bibfnamefont {G.}~\bibnamefont {Ceder}}, \bibinfo {author} {\bibfnamefont {M.}~\bibnamefont {Asta}},\ and\ \bibinfo {author} {\bibfnamefont {P.~D.}\ \bibnamefont {Tepesch}},\ }\bibfield  {title} {\bibinfo {title} {First-principles theory of ionic diffusion with nondilute carriers},\ }\href {https://doi.org/10.1103/PhysRevB.64.184307} {\bibfield  {journal} {\bibinfo  {journal} {Physical Review B}\ }\textbf {\bibinfo {volume} {64}},\ \bibinfo {pages} {184307} (\bibinfo {year} {2001})}\BibitemShut {NoStop}%
\bibitem [{\citenamefont {Natarajan}\ and\ \citenamefont {Van~der Ven}(2018)}]{natarajan_machine-learning_2018}%
  \BibitemOpen
  \bibfield  {author} {\bibinfo {author} {\bibfnamefont {A.~R.}\ \bibnamefont {Natarajan}}\ and\ \bibinfo {author} {\bibfnamefont {A.}~\bibnamefont {Van~der Ven}},\ }\bibfield  {title} {{\bibinfo {title} {Machine-learning the configurational energy of multicomponent crystalline solids}},\ }\href {https://doi.org/10.1038/s41524-018-0110-y} {\bibfield  {journal} {\bibinfo  {journal} {npj Computational Materials}\ }\textbf {\bibinfo {volume} {4}},\ \bibinfo {pages} {56} (\bibinfo {year} {2018})}\BibitemShut {NoStop}%
\bibitem [{\citenamefont {Müller}\ and\ \citenamefont {Natarajan}(2025)}]{muller_constructing_2025}%
  \BibitemOpen
  \bibfield  {author} {\bibinfo {author} {\bibfnamefont {Y.~L.}\ \bibnamefont {Müller}}\ and\ \bibinfo {author} {\bibfnamefont {A.~R.}\ \bibnamefont {Natarajan}},\ }\bibfield  {title} {\bibinfo {title} {Constructing multicomponent cluster expansions with machine-learning and chemical embedding},\ }\href {https://doi.org/10.1038/s41524-025-01543-3} {\bibfield  {journal} {\bibinfo  {journal} {npj Computational Materials}\ }\textbf {\bibinfo {volume} {11}},\ \bibinfo {pages} {1} (\bibinfo {year} {2025})}\BibitemShut {NoStop}%
\bibitem [{\citenamefont {Darken}(1948)}]{darken_diffusion_1948}%
  \BibitemOpen
  \bibfield  {author} {\bibinfo {author} {\bibfnamefont {L.~S.}\ \bibnamefont {Darken}},\ }\bibfield  {title} {\bibinfo {title} {Diffusion, {Mobility} and {Their} {Interrelation} through {Free} {Energy} in {Binary} {Metallic} {Systems}},\ }\href@noop {} {\bibfield  {journal} {\bibinfo  {journal} {Transactions of the Metallurgical Society of AIME}\ }\textbf {\bibinfo {volume} {175}},\ \bibinfo {pages} {184} (\bibinfo {year} {1948})}\BibitemShut {NoStop}%
\bibitem [{\citenamefont {Van~der Ven}\ \emph {et~al.}(2012)\citenamefont {Van~der Ven}, \citenamefont {Bhattacharya},\ and\ \citenamefont {Belak}}]{van_der_ven_understanding_2012}%
  \BibitemOpen
  \bibfield  {author} {\bibinfo {author} {\bibfnamefont {A.}~\bibnamefont {Van~der Ven}}, \bibinfo {author} {\bibfnamefont {J.}~\bibnamefont {Bhattacharya}},\ and\ \bibinfo {author} {\bibfnamefont {A.~A.}\ \bibnamefont {Belak}},\ }\bibfield  {title} {\bibinfo {title} {Understanding {Li} {Diffusion} in {Li}-{Intercalation} {Compounds}},\ }\href {https://doi.org/10.1021/ar200329r} {\bibfield  {journal} {\bibinfo  {journal} {Accounts of Chemical Research}\ }\textbf {\bibinfo {volume} {46}},\ \bibinfo {pages} {1216} (\bibinfo {year} {2012})}\BibitemShut {NoStop}%
\bibitem [{\citenamefont {Kolli}\ and\ \citenamefont {Van~der Ven}(2021)}]{kolli_elucidating_2021}%
  \BibitemOpen
  \bibfield  {author} {\bibinfo {author} {\bibfnamefont {S.~K.}\ \bibnamefont {Kolli}}\ and\ \bibinfo {author} {\bibfnamefont {A.}~\bibnamefont {Van~der Ven}},\ }\bibfield  {title} {\bibinfo {title} {Elucidating the {Factors} {That} {Cause} {Cation} {Diffusion} {Shutdown} in {Spinel}-{Based} {Electrodes}},\ }\href {https://doi.org/10.1021/acs.chemmater.1c01668} {\bibfield  {journal} {\bibinfo  {journal} {Chemistry of Materials}\ }\textbf {\bibinfo {volume} {33}},\ \bibinfo {pages} {6421} (\bibinfo {year} {2021})}\BibitemShut {NoStop}%
\bibitem [{\citenamefont {Senkov}\ \emph {et~al.}(2011)\citenamefont {Senkov}, \citenamefont {Wilks}, \citenamefont {Scott},\ and\ \citenamefont {Miracle}}]{senkov_mechanical_2011}%
  \BibitemOpen
  \bibfield  {author} {\bibinfo {author} {\bibfnamefont {O.~N.}\ \bibnamefont {Senkov}}, \bibinfo {author} {\bibfnamefont {G.~B.}\ \bibnamefont {Wilks}}, \bibinfo {author} {\bibfnamefont {J.~M.}\ \bibnamefont {Scott}},\ and\ \bibinfo {author} {\bibfnamefont {D.~B.}\ \bibnamefont {Miracle}},\ }\bibfield  {title} {\bibinfo {title} {Mechanical properties of {Nb25Mo25Ta25W25} and {V20Nb20Mo20Ta20W20} refractory high entropy alloys},\ }\href {https://doi.org/10.1016/j.intermet.2011.01.004} {\bibfield  {journal} {\bibinfo  {journal} {Intermetallics}\ }\textbf {\bibinfo {volume} {19}},\ \bibinfo {pages} {698} (\bibinfo {year} {2011})}\BibitemShut {NoStop}%
\bibitem [{\citenamefont {Shang}\ \emph {et~al.}(2016)\citenamefont {Shang}, \citenamefont {Zhou}, \citenamefont {Wang}, \citenamefont {Ross}, \citenamefont {Liu}, \citenamefont {Hu}, \citenamefont {Fang}, \citenamefont {Wang},\ and\ \citenamefont {Liu}}]{shang_comprehensive_2016}%
  \BibitemOpen
  \bibfield  {author} {\bibinfo {author} {\bibfnamefont {S.-L.}\ \bibnamefont {Shang}}, \bibinfo {author} {\bibfnamefont {B.-C.}\ \bibnamefont {Zhou}}, \bibinfo {author} {\bibfnamefont {W.~Y.}\ \bibnamefont {Wang}}, \bibinfo {author} {\bibfnamefont {A.~J.}\ \bibnamefont {Ross}}, \bibinfo {author} {\bibfnamefont {X.~L.}\ \bibnamefont {Liu}}, \bibinfo {author} {\bibfnamefont {Y.-J.}\ \bibnamefont {Hu}}, \bibinfo {author} {\bibfnamefont {H.-Z.}\ \bibnamefont {Fang}}, \bibinfo {author} {\bibfnamefont {Y.}~\bibnamefont {Wang}},\ and\ \bibinfo {author} {\bibfnamefont {Z.-K.}\ \bibnamefont {Liu}},\ }\bibfield  {title} {\bibinfo {title} {A comprehensive first-principles study of pure elements: {Vacancy} formation and migration energies and self-diffusion coefficients},\ }\href {https://doi.org/10.1016/j.actamat.2016.02.031} {\bibfield  {journal} {\bibinfo  {journal} {Acta Materialia}\ }\textbf {\bibinfo {volume} {109}},\ \bibinfo {pages} {128} (\bibinfo {year} {2016})}\BibitemShut {NoStop}%
\bibitem [{\citenamefont {Van~der Ven}\ and\ \citenamefont {Ceder}(2005)}]{van_der_ven_first_2005}%
  \BibitemOpen
  \bibfield  {author} {\bibinfo {author} {\bibfnamefont {A.}~\bibnamefont {Van~der Ven}}\ and\ \bibinfo {author} {\bibfnamefont {G.}~\bibnamefont {Ceder}},\ }\bibfield  {title} {\bibinfo {title} {First {Principles} {Calculation} of the {Interdiffusion} {Coefficient} in {Binary} {Alloys}},\ }\href {https://doi.org/10.1103/PhysRevLett.94.045901} {\bibfield  {journal} {\bibinfo  {journal} {Physical Review Letters}\ }\textbf {\bibinfo {volume} {94}},\ \bibinfo {pages} {045901} (\bibinfo {year} {2005})}\BibitemShut {NoStop}%
\bibitem [{\citenamefont {Behara}\ and\ \citenamefont {Van~der Ven}(2024)}]{behara_role_2024}%
  \BibitemOpen
  \bibfield  {author} {\bibinfo {author} {\bibfnamefont {S.~S.}\ \bibnamefont {Behara}}\ and\ \bibinfo {author} {\bibfnamefont {A.}~\bibnamefont {Van~der Ven}},\ }\bibfield  {title} {\bibinfo {title} {Role of {Short}-{Range} {Order} on {Diffusion} {Coefficients} in the {Li}–{Mg} {Alloy}},\ }\href {https://doi.org/10.1021/acs.chemmater.4c02334} {\bibfield  {journal} {\bibinfo  {journal} {Chemistry of Materials}\ }\textbf {\bibinfo {volume} {36}},\ \bibinfo {pages} {11236} (\bibinfo {year} {2024})}\BibitemShut {NoStop}%
\bibitem [{\citenamefont {Mishin}\ and\ \citenamefont {Farkas}(1997{\natexlab{c}})}]{mishin_monte_1997}%
  \BibitemOpen
  \bibfield  {author} {\bibinfo {author} {\bibfnamefont {Y.}~\bibnamefont {Mishin}}\ and\ \bibinfo {author} {\bibfnamefont {D.}~\bibnamefont {Farkas}},\ }\bibfield  {title} {\bibinfo {title} {Monte {Carlo} simulation of correlation effects in a random bcc alloy},\ }\href {https://doi.org/10.1080/01418619708210291} {\bibfield  {journal} {\bibinfo  {journal} {Philosophical Magazine A}\ }\textbf {\bibinfo {volume} {75}},\ \bibinfo {pages} {201} (\bibinfo {year} {1997}{\natexlab{c}})}\BibitemShut {NoStop}%
\bibitem [{\citenamefont {Yu}\ \emph {et~al.}(2008)\citenamefont {Yu}, \citenamefont {Van~der Ven},\ and\ \citenamefont {Thornton}}]{yu_theory_2008}%
  \BibitemOpen
  \bibfield  {author} {\bibinfo {author} {\bibfnamefont {H.-C.}\ \bibnamefont {Yu}}, \bibinfo {author} {\bibfnamefont {A.}~\bibnamefont {Van~der Ven}},\ and\ \bibinfo {author} {\bibfnamefont {K.}~\bibnamefont {Thornton}},\ }\bibfield  {title} {\bibinfo {title} {Theory of grain boundary diffusion induced by the {Kirkendall} effect},\ }\href {https://doi.org/10.1063/1.2978161} {\bibfield  {journal} {\bibinfo  {journal} {Applied Physics Letters}\ }\textbf {\bibinfo {volume} {93}},\ \bibinfo {pages} {091908} (\bibinfo {year} {2008})}\BibitemShut {NoStop}%
\bibitem [{\citenamefont {Soisson}\ and\ \citenamefont {Fu}(2007)}]{soisson_cu-precipitation_2007}%
  \BibitemOpen
  \bibfield  {author} {\bibinfo {author} {\bibfnamefont {F.}~\bibnamefont {Soisson}}\ and\ \bibinfo {author} {\bibfnamefont {C.-C.}\ \bibnamefont {Fu}},\ }\bibfield  {title} {\bibinfo {title} {Cu-precipitation kinetics in $\alpha$-{Fe} from atomistic simulations: {Vacancy}-trapping effects and {Cu}-cluster mobility},\ }\href {https://doi.org/10.1103/PhysRevB.76.214102} {\bibfield  {journal} {\bibinfo  {journal} {Physical Review B}\ }\textbf {\bibinfo {volume} {76}},\ \bibinfo {pages} {214102} (\bibinfo {year} {2007})}\BibitemShut {NoStop}%
\bibitem [{\citenamefont {Nastar}\ and\ \citenamefont {Soisson}(2012)}]{nastar_atomistic_2012}%
  \BibitemOpen
  \bibfield  {author} {\bibinfo {author} {\bibfnamefont {M.}~\bibnamefont {Nastar}}\ and\ \bibinfo {author} {\bibfnamefont {F.}~\bibnamefont {Soisson}},\ }\bibfield  {title} {\bibinfo {title} {Atomistic modeling of phase transformations: {Point}-defect concentrations and the time-scale problem},\ }\href {https://doi.org/10.1103/PhysRevB.86.220102} {\bibfield  {journal} {\bibinfo  {journal} {Physical Review B}\ }\textbf {\bibinfo {volume} {86}},\ \bibinfo {pages} {220102} (\bibinfo {year} {2012})}\BibitemShut {NoStop}%
\bibitem [{\citenamefont {Li}\ \emph {et~al.}(2024)\citenamefont {Li}, \citenamefont {Schuler}, \citenamefont {Fu},\ and\ \citenamefont {Nastar}}]{li_vacancy_2024}%
  \BibitemOpen
  \bibfield  {author} {\bibinfo {author} {\bibfnamefont {K.}~\bibnamefont {Li}}, \bibinfo {author} {\bibfnamefont {T.}~\bibnamefont {Schuler}}, \bibinfo {author} {\bibfnamefont {C.-C.}\ \bibnamefont {Fu}},\ and\ \bibinfo {author} {\bibfnamefont {M.}~\bibnamefont {Nastar}},\ }\bibfield  {title} {\bibinfo {title} {Vacancy formation free energy in concentrated alloys: {Equilibrium} vs. random sampling},\ }\href {https://doi.org/10.1016/j.actamat.2024.120355} {\bibfield  {journal} {\bibinfo  {journal} {Acta Materialia}\ }\textbf {\bibinfo {volume} {281}},\ \bibinfo {pages} {120355} (\bibinfo {year} {2024})}\BibitemShut {NoStop}%
\end{thebibliography}%

\appendix

\crefalias{section}{appendix}
\section{Eigenspectra for a binary alloy}
\label{sec-app:binary-alloy}

This appendix derives the transformed thermodynamic factor and the eigenvalues and eigenvectors of the binary diffusion matrix quoted in \cref{sec:coord-transf-flux}. Applying \cref{eq:thermodynamic_factor_transformation} to \cref{eq:binary_thermodynamic_factor_matrix_activity} gives the transformed thermodynamic factor:
\begin{widetext}
\begin{equation}
  \label{eq:transformed_binary_thermodynamic_factor}
  \mathbf{\tilde{\Theta}}_{\mathbf{y}} =
    \begin{bmatrix}
    \frac{4}{x_{\mathrm{Va}}}+\frac{1}{x_{1}}+\frac{1}{x_{2}} +(f_{11}+2f_{12}+f_{22}) & \frac{1}{x_{1}}-\frac{1}{x_{2}}+f_{11}-f_{22}\\ \frac{1}{x_{1}}-\frac{1}{x_{2}}+f_{11}-f_{22} & \frac{1}{x_{1}}+\frac{1}{x_{2}} + (f_{11}+f_{22}-2f_{12})
  \end{bmatrix} = \frac{1}{k_\mathrm{B}T}
  \begin{bmatrix}
    \frac{\partial^{2}g}{\partial y_{1}^{2}} & \frac{\partial^{2}g}{\partial y_{1}\partial y_{2}} \\ \frac{\partial^{2}g}{\partial y_{2}\partial y_{1}} & \frac{\partial^{2}g}{\partial y_{2}^{2}}
  \end{bmatrix}
\end{equation}
\end{widetext}

\noindent The $1/x_{\mathrm{Va}}$ terms of \cref{eq:binary_thermodynamic_factor_matrix_activity} cancel in every element except the first, where they combine into $4/x_{\mathrm{Va}}$. Collecting the terms that remain in the first row into the two coefficients
\begin{align}
  a &= \frac{1}{x_{1}}+\frac{1}{x_{2}}+f_{11}+2f_{12}+f_{22}
  \label{eq:theta_y_a} \\
  b &= \frac{1}{x_{1}}-\frac{1}{x_{2}}+f_{11}-f_{22}
  \label{eq:theta_y_b}
\end{align}
\noindent recovers the compact form of \cref{eq:transformed_binary_thermodynamic_factor_compact}.

Multiplying the transformed transport coefficients of \cref{eq:transformed_mobility} by the transformed thermodynamic factor of \cref{eq:transformed_binary_thermodynamic_factor_compact} gives the diffusion matrix of the binary alloy in the $\mathbf{y}$ coordinates:
\begin{widetext}
\begin{equation}
  \label{eq:transformed_diffusion_matrix}
  \mathbf{D}_{\mathbf{y}} = \frac{1}{4}
  \begin{bmatrix}
    4D^{\star}_{\mathrm{Va}} + a D^{\star}_{\mathrm{Va}}x_{\mathrm{Va}} + b\left(\tilde{L}_{11}-\tilde{L}_{22}\right) &
    b D^{\star}_{\mathrm{Va}}x_{\mathrm{Va}} + \Theta_{\mathrm{mix}}\left(\tilde{L}_{11}-\tilde{L}_{22}\right) \\
    \left(\frac{4}{x_{\mathrm{Va}}}+a\right)\left(\tilde{L}_{11}-\tilde{L}_{22}\right) + b\left(\tilde{L}_{11}+\tilde{L}_{22}-2\tilde{L}_{12}\right) &
    b\left(\tilde{L}_{11}-\tilde{L}_{22}\right) + \Theta_{\mathrm{mix}}\left(\tilde{L}_{11}+\tilde{L}_{22}-2\tilde{L}_{12}\right)
  \end{bmatrix}
\end{equation}
\end{widetext}
The coefficient $a$ appears only in the two elements of the first column, and in both it multiplies a quantity already proportional to the vacancy concentration. The eigenvalues follow most easily from the trace and the determinant of \cref{eq:transformed_diffusion_matrix}. \Citeauthor{van_der_ven_vacancy_2010} \cite{van_der_ven_vacancy_2010} provides a detailed derivation.

Denoting the elements of \cref{eq:transformed_diffusion_matrix} by $D_{\mathbf{y},ij}$, the eigenvectors follow from the second row of $\left(\mathbf{D}_{\mathbf{y}}-\lambda\mathbf{I}\right)\mathbf{v}=\mathbf{0}$ as
\begin{equation}
  \label{eq:eigenvectors_binary}
  \mathbf{v}_{\pm} \propto
  \begin{bmatrix}
    \left(D_{\mathbf{y},11}-D_{\mathbf{y},22}\right) \pm \sqrt{\left(D_{\mathbf{y},11}-D_{\mathbf{y},22}\right)^{2}+4D_{\mathbf{y},12}D_{\mathbf{y},21}} \\
    2D_{\mathbf{y},21}
  \end{bmatrix}
\end{equation}
where the difference of the diagonal elements simplifies to $D_{\mathbf{y},11}-D_{\mathbf{y},22} = D^{\star}_{\mathrm{Va}} + \left[a D^{\star}_{\mathrm{Va}}x_{\mathrm{Va}} - \Theta_{\mathrm{mix}}(\tilde{L}_{11}+\tilde{L}_{22}-2\tilde{L}_{12})\right]/4$, since the terms in $b$ cancel. Substituting the elements of \cref{eq:transformed_diffusion_matrix} into \cref{eq:eigenvectors_binary} and expanding the square root for small $x_{\mathrm{Va}}$ gives the limiting eigenvectors of \cref{eq:eigenvector_binary_plus,eq:eigenvector_binary_minus}. The expansion uses the proportionality of every $\tilde{L}_{ij}$ to the vacancy concentration.

\section{Multicomponent alloy thermodynamic factor projection}
\label{sec-app:multicomponent-alloy-projection}

Define $\hat{\mu}_{i} = \mu_{i} - \mu_{1}$ for all $i\neq 1$. The corresponding characteristic potential $\hat{G}(x_{2},x_{3},\cdots,x_{c},x_{\Va},M,T)$ has as its natural variables the compositions of every species other than species 1, whose composition follows from $x_{1} = 1 - x_{\Va} - \sum_{k=2}^{c}x_{k}$. The second derivatives of $\hat{G}$ at fixed total number of sites $M$ and temperature $T$ are then
\begin{widetext}
\begin{align}
  \label{eq:projected_thermodynamic_factor_proof}
  \left(\frac{\partial \hat{\mu}_{i}}{\partial x_{j}}\right)_{x_{k\neq j,1},x_{\Va}}
  = \left(\frac{\partial \hat{\mu}_{i}}{\partial x_{j}}\right)_{x_{k\neq j}} + \left(\frac{\partial \hat{\mu}_{i}}{\partial x_{1}}\right)_{x_{k\neq 1}} \left(\frac{\partial x_{1}}{\partial x_{j}}\right)_{x_{k\neq j,1},x_{\Va}}
  = \left(\frac{\partial \hat{\mu}_{i}}{\partial x_{j}}\right)_{x_{k\neq j}} - \left(\frac{\partial \hat{\mu}_{i}}{\partial x_{1}}\right)_{x_{k\neq 1}}
\end{align}
\end{widetext}

\noindent where the second equality uses $\left(\partial x_{1}/\partial x_{j}\right)_{x_{k\neq j,1},x_{\Va}} = -1$, which holds because the elemental compositions sum to $1-x_{\Va}$ at fixed vacancy concentration.

Collecting these derivatives for the $c-1$ compositions that exclude the vacancy and species 1 gives a square matrix of dimension $(c-1)\times (c-1)$, which we denote $\mathbf{\Theta}_{\mathrm{alloy}} = k_\mathrm{B}T \mathbf{\tilde{\Theta}}_{\mathrm{alloy}}$. The subscript indicates that this thermodynamic factor reduces to that of the vacancy-free alloy when evaluated at $x_{\Va}=0$.

\section{Eigenspectrum of $\mathbf{D}$ in the dilute-vacancy limit}
\label{sec-app:depend-eigensp-diff}

This appendix constructs the transformation that decouples vacancy transport from alloy interdiffusion in the dilute-vacancy limit, and uses it to show that the off-diagonal blocks of the transformed thermodynamic factor reach the eigenvalues of the diffusion matrix only at second order in the vacancy concentration, so that discarding them in \cref{eq:general_approximate_transformed_thermodynamic_factor} leaves the eigenspectrum unchanged to leading order. Separating the divergent element from the rest gives
\begin{equation}
  \label{eq:thermodynamic_factor_partition}
  \mathbf{\tilde{\Theta}}_{\mathbf{y}} = \frac{c^{2}}{x_{\Va}}\vec{e}_{1} \vec{e}_{1}^{\mathrm{T}} +
  \begin{bmatrix}
    a & \vec{b}^{\mathrm{T}} \\
    \vec{b} & \mathbf{\tilde{\Theta}}_{\mathrm{alloy}}
  \end{bmatrix} = \frac{c^{2}}{x_{\Va}}\vec{e}_{1} \vec{e}_{1}^{\mathrm{T}} + \mathbf{S}
\end{equation}
where $\vec{e}_{1} = [1,0,\cdots,0]^{\mathrm{T}}$ is a vector of dimension $c$. The scalar $a$ and the vector $\vec{b}$ play the same role as $a$ and $b$ in \cref{eq:theta_y_a,eq:theta_y_b}, although their expressions are more involved. Neither contains $x_{\Va}$, and both retain the interactions of the vacancy with the alloy. The transport coefficients are rescaled in the same way
\begin{equation}
  \label{eq:scaled_mobility}
  \mathbf{\tilde{L}}_{\mathbf{y}} = x_{\Va} \boldsymbol{\Lambda}
\end{equation}
where $\boldsymbol{\Lambda}$ is a square matrix independent of the vacancy concentration. The matrix of diffusion coefficients is then
\begin{equation}
  \label{eq:transformed_diffusion_constant}
  \mathbf{D}_{\mathbf{y}} = c^{2} \boldsymbol{\Lambda} \vec{e}_{1}\vec{e}_{1}^{\mathrm{T}} + x_{\Va} \boldsymbol{\Lambda} \mathbf{S} = \mathbf{D}_{0} + x_{\Va}\mathbf{D}_{1}
\end{equation}

Define $\vec{u}=\boldsymbol{\Lambda}\vec{e}_{1}$, which is a right eigenvector of $\mathbf{D}_{0}$ with an eigenvalue of exactly $D^{\star}_{\Va}$, and let $\mathbf{V} = [\vec{e}_{2},\cdots \vec{e}_{c}]$, where $\vec{e}_{i}$ is the unit vector of dimension $c$ whose entries all vanish except the $i^{\mathrm{th}}$, which equals one. The columns of $\mathbf{V}$ span the right null space of $\mathbf{D}_{0}$. Define $\mathbf{P} = [\vec{u}, \mathbf{V}]$ and its inverse $\mathbf{P}^{-1} = [\vec{p},\mathbf{R}]^{\mathrm{T}}$, which requires $\vec{p}^{\mathrm{T}}\vec{u} = 1$, $\vec{p}^{\mathrm{T}}\mathbf{V} = \vec{0}^{\mathrm{T}}$, $\mathbf{R}^{T}\vec{u} =  \vec{0}$, and $\mathbf{R}^{T}\mathbf{V} = \mathbf{I}$. We define another compositional coordinate transformation as
\begin{align}
    \mathbf{z} = \mathbf{P}^{-1}\mathbf{y}.
    \label{eq:comp_trans_2}
\end{align}
Transforming $\mathbf{D}_{\mathbf{y}}$ to the $\mathbf{z}-$coordinate space leaves the eigenvalues unchanged and gives
\begin{equation}
  \label{eq:transformation_D}
\mathbf{D}_{\mathbf{z}} =  \mathbf{P}^{-1}\mathbf{D}_{\mathbf{y}} \mathbf{P} =
  \begin{bmatrix}
    D^{\star}_{\Va} + x_{\Va} b_{11} & x_{\Va} \vec{b}_{12}^{\mathrm{T}} \\
    x_{\Va} \vec{b}_{21} & x_{\Va} \mathbf{B}
  \end{bmatrix}
\end{equation}

\noindent The coefficients $a$ and $\vec{b}$ of \cref{eq:thermodynamic_factor_partition} reach the transformed diffusion matrix only through $b_{11}$, $\vec{b}_{12}$ and $\vec{b}_{21}$, and every term that depends on the vacancy concentration carries an explicit factor of $x_{\Va}$. The $(c-1)\times (c-1)$ block in the lower right of \cref{eq:transformation_D} contains no contribution from $a$ or $\vec{b}$ and is exactly
\begin{equation}
  \label{eq:B_matrix_in_transformed_D}
  \mathbf{B} = \mathbf{R}^{T} \boldsymbol{\Lambda} \mathbf{V} \mathbf{\tilde{\Theta}}_{\mathrm{alloy}}
\end{equation}

\Cref{eq:transformation_D} shows that $\mathbf{P} = [\vec{u},\mathbf{V}]$ block diagonalizes the diffusion matrix in the dilute-vacancy limit. Collecting the terms that survive at $x_{\Va}=0$ in each block,
\begin{equation}
  \label{eq:block_diagonal_form}
  \mathbf{D}_{\mathbf{z}} =
  \begin{bmatrix}
    D^{\star}_{\Va} & \vec{0}^{\mathrm{T}} \\
    \vec{0} & x_{\Va}\mathbf{B}
  \end{bmatrix}
  + x_{\Va}
  \begin{bmatrix}
    b_{11} & \vec{b}_{12}^{\mathrm{T}} \\
    \vec{b}_{21} & \mathbf{0}
  \end{bmatrix}
\end{equation}

\noindent which is exact. The first column of $\mathbf{P}$, $\vec{u} = \boldsymbol{\Lambda}\vec{e}_{1}$, spans the fast mode, whose rate is the vacancy tracer diffusion coefficient and which carries no dependence on the thermodynamics of the alloy. The columns of $\mathbf{V}$ span the vacancy-free composition hyperplane, within which the $c-1$ slow modes are governed by $\mathbf{B}$ and therefore by the free-energy curvature of the vacancy-free alloy alone. The two blocks are separated by a factor of $D^{\star}_{\Va}/x_{\Va}$ in magnitude. The second term of \cref{eq:block_diagonal_form} contains the contributions of $a$ and $\vec{b}$. It modifies the fast eigenvalue at relative order $x_{\Va}$, and the slow block only through the Schur complement of \cref{eq:transformed_char_equation_schur_complement}, where the separation of the two blocks suppresses it by a further factor of the vacancy concentration. The remainder of this appendix establishes both statements, for the eigenvalues and then for the eigenvectors.

The eigenvalues of \cref{eq:transformation_D} are identical to those of $\mathbf{D}_{\mathbf{x}}$, which are the ones sought, so it suffices to work with \cref{eq:transformation_D}. Substituting $\lambda = x_{\Va}\kappa$ for an eigenvalue, which writes the scaling with vacancy concentration out explicitly, the characteristic equation reads
\begin{equation}
  \label{eq:transformed_char_equation}
  \det
  \begin{bmatrix}
    D^{\star}_{\Va} + x_{\Va} (b_{11} - \kappa) & x_{\Va} \vec{b}_{12}^{\mathrm{T}} \\
    x_{\Va} \vec{b}_{21} & x_{\Va} (\mathbf{B} - \kappa \mathbf{I}_{(c-1)})
  \end{bmatrix}
  = 0
\end{equation}

Using the Schur complement, and dividing out an overall factor of $x_{\Va}^{c-1}$ that cannot vanish, solving \cref{eq:transformed_char_equation} for $\kappa$ is equivalent to solving
\begin{widetext}
\begin{equation}
  \label{eq:transformed_char_equation_schur_complement}
  \left(D^{\star}_{\Va} + x_{\Va}(b_{11}-\kappa)\right) \det\left( (\mathbf{B} - \kappa \mathbf{I}_{(c-1)}) - \frac{x_{\Va}\vec{b}_{21}\vec{b}_{12}^{\mathrm{T}}}{D_{\Va}^{\star}+x_{\Va}(b_{11}-\kappa)}\right)  = 0
\end{equation}
\end{widetext}
The factorization requires the prefactor to be non-zero. At any root for which $\kappa$ stays finite the prefactor tends to $D^{\star}_{\Va}$, so the determinant supplies the $c-1$ slow eigenvalues. The one remaining eigenvalue is the root along which $\kappa$ diverges as the vacancies become dilute. Setting the prefactor to zero locates it at $x_{\Va}\kappa_{1} = D^{\star}_{\Va}+x_{\Va}b_{11}$, so that the first and largest eigenvalue is $\lambda_{1} \approx D^{\star}_{\Va}$ to leading order in the vacancy concentration. The rescaling by $x_{\Va}$ is therefore useful only for the $c-1$ eigenvalues that vanish with the vacancy concentration.

The remaining $c-1$ eigenvalues follow from the determinant. Denote the matrix whose determinant appears in \cref{eq:transformed_char_equation_schur_complement} by $\mathbf{T}$, so that the condition reads $F(\kappa,x_{\Va}) = \det(\mathbf{T}) = 0$, with $\mathbf{T}$ a function of both $\kappa$ and $x_{\Va}$. Its solutions $\kappa$ are therefore themselves functions of $x_{\Va}$. Each of these $c-1$ solutions stays finite as the vacancies become dilute and can be Taylor expanded in the vacancy concentration as
\begin{equation}
  \label{eq:eigenvalue_taylor_expansion}
  \kappa_{i}(x_{\Va}) = \kappa_{i}^{0} + x_{\Va} \kappa_{i}^{\prime}(x_{\Va})\vert_{x_{\Va}=0} + \mathcal{O}(x_{\Va}^{2})
\end{equation}
where $\kappa_{i}^{0}$ are the eigenvalues of $\mathbf{B}$, obtained by solving $\det\left(\mathbf{B}-\kappa\mathbf{I}_{(c-1)}\right)=0$, and equivalently the roots of $F(\kappa,0)=0$ at zero vacancy concentration. The next term in \cref{eq:eigenvalue_taylor_expansion} is linear in the vacancy concentration and holds the derivative of the eigenvalue with respect to it, which requires that derivative to exist and be finite. That derivative is also where the contributions from $b_{11}$, $\vec{b}_{12}$ and $\vec{b}_{21}$ appear. Since the eigenvalues sought are the $\lambda_{i}$, the remaining $c-1$ eigenvalues of the diffusion matrix are
\begin{equation}
  \label{eq:eigenvalue_final_taylor}
  \lambda_{i} = x_{\Va} \kappa_{i} = x_{\Va} \kappa_{i}^{0} + \mathcal{O}(x_{\Va}^{2})
\end{equation}

\noindent so the dependence on $b_{11}$, $\vec{b}_{12}$ and $\vec{b}_{21}$ lies entirely within the term quadratic in the vacancy concentration. Setting these terms to zero therefore induces an error in the eigenvalues of order $x_{\Va}^{2}$, which vanishes as $x_{\Va}\rightarrow 0$. 

Let the eigenvectors of $\mathbf{D}_{\mathbf{z}}$ be partitioned as $\vec{v}_{\mathbf{z}} = [v_u, \vec{v}^{\mathrm{T}}_\mathrm{alloy} ]^\mathrm{T}$, where $\vec{v}_\mathrm{alloy}$ has $c-1$ components. With $\mathbf{D}_{\mathbf{z}}$ given by \cref{eq:transformation_D} and the eigenvalue written as $\lambda = x_{\Va}\kappa$, as in \cref{eq:eigenvalue_final_taylor}, the two components satisfy
\begin{align}
\label{eq:eigenvecs_M_1}
    &\left(D^{\star}_{\Va} + x_{\Va} b_{11}\right) v_u + x_\Va \vec{b}_{12}^\mathrm{T} \vec{v}_\mathrm{alloy} = x_\Va \kappa v_u   \\
&x_\Va \vec{b}_{21} v_u + x_\Va \mathbf{B} \vec{v}_\mathrm{alloy} = x_\Va \kappa \vec{v}_\mathrm{alloy}.
\label{eq:eigenvecs_M_2}
\end{align}
The $c-1$ slow eigenvalues scale linearly with $x_{\Va}$ to leading order, so $\kappa$ remains finite and \cref{eq:eigenvecs_M_1} forces $v_{u}$ to be of order $x_{\Va}$ for the corresponding eigenvectors. Dividing \cref{eq:eigenvecs_M_2} by $x_{\Va}$ and using this scaling of $v_{u}$ reduces it to the eigenvalue problem of $\mathbf{B}$, so that
\begin{align}
    \vec{v}_\mathrm{alloy} =  \vec{v}^0_\mathrm{alloy}  + \mathcal{O}(x_\Va),
\end{align}
where $\vec{v}^0_\mathrm{alloy}$ are the $c-1$ dimensional eigenvectors of $\mathbf{B}$. For the first eigenvector, $x_{\Va}\kappa_{1} = D^{\star}_{\Va}+x_{\Va}b_{11}$ turns the right hand side of \cref{eq:eigenvecs_M_2} into $\left(D^{\star}_{\Va}+x_{\Va}b_{11}\right)\vec{v}_\mathrm{alloy}$, whose leading term $D^{\star}_{\Va} \vec{v}_\mathrm{alloy}$ has no counterpart of the same order on the left, and therefore $\vec{v}_\mathrm{alloy} = \vec{0}$ to leading order. The eigenvectors of $\mathbf{D}_{\mathbf{z}}$ are then
\begin{align}
\vec{v}_{\mathbf{z},1} &= [1,\vec{0}^\mathrm{T}]^\mathrm{T} + \mathcal{O}(x_\Va) \quad \text{and} \nonumber \\
\vec{v}_{\mathbf{z},1+i} &= [0, \vec{v}_{\mathrm{alloy}_i}^\mathrm{T} ]^\mathrm{T} + \mathcal{O}(x_\Va)
\label{eq:eigenvec_solutions_M}
\end{align}
for $i \in \{1,2\ldots c-1\}$. As with the eigenvalues, the dependence of $\vec{v}_{\mathbf{z},1}$ on $\vec{b}_{21}$, and of the remaining eigenvectors on $\vec{b}_{12}$, $\vec{b}_{21}$ and $b_{11}$, is confined to corrections linear in the vacancy concentration, which vanish as $x_\Va \rightarrow 0$.

The eigenvectors in the $\mathbf{y}$ coordinates follow as $\vec{v}_{\mathbf{y}} = \mathbf{P} \vec{v}_{\mathbf{z}}$.

\section{Derivation of the single-vacancy covariance matrix}
\label{sec-app:cg_derivation}

This derivation closely follows \cite{belak_effect_2015,lee_modeling_2026}, and only the salient steps are reproduced here.

The semi-grand potential of a material with $c$ elements and vacancies distributed over $M$ sites at a temperature $T$ is given by:
\begin{equation}
	\label{eq:semi_grand_potential}
	\Omega(\vec{\sigma}^\prime) = E(\vec{\sigma}^\prime) - \sum_{i=1}^c \tilde{\mu}_iN_i(\vec{\sigma}^\prime)
\end{equation}
where $E$ is the total energy of a configuration $\vec{\sigma}^\prime$ of the $c$ elements and vacancies over the $M$ sites. The corresponding semi-grand canonical partition function is:
\begin{equation}
	\label{eq:semi_grand_partition_function}
	Z = \sum_{\vec{\sigma}^\prime} \exp(-\beta \Omega(\vec{\sigma}^\prime))
\end{equation}
where $\beta = 1/(k_\mathrm{B} T)$ and the sum runs over all possible arrangements of the $c$ elements and vacancies over the $M$ sites. In the dilute-vacancy limit, \cref{eq:semi_grand_partition_function} can be decomposed using the vacancy partition function, the partition function of the vacancy-free alloy, and an auxiliary quantity $\xi$:
\begin{align}
	\label{eq:vacancy_partition_function}
	Z_\mathrm{Va}(\vec{\sigma},\mu_\mathrm{Va}) & = \frac{1}{c}\sum_{\vec{\nu}} \exp\left(-\beta (E(\vec{\nu}) - E(\vec{\sigma}) -\mu_\mathrm{Va} + \mu_{f(\vec{\nu})})\right)
	\nonumber \\
    & =\exp{\left(\beta \mu_\mathrm{Va}\right)} Z_\mathrm{Va}(\vec{\sigma})
\end{align}

\begin{equation}
	\label{eq:alloy_partition_function}
	\tilde{Z} = \sum_{\vec{\sigma}} \exp{\left( -\beta \Omega(\vec{\sigma}) \right)}
\end{equation}
\begin{align}
	\label{eq:xi}
	\xi(\mu_\mathrm{Va}) & = \sum_{\vec{\sigma}} \frac{\exp{\left(-\beta \Omega(\vec{\sigma})\right)}}{\tilde{Z}} Z_\mathrm{Va}(\vec{\sigma},\mu_\mathrm{Va}) \\
	& = \exp(\beta \mu_\mathrm{Va})\langle Z_\mathrm{Va}\rangle_{\textrm{alloy}}
\end{align}
The sum over $\vec{\sigma}$ runs over all chemical arrangements of the $c$ elements over the $M$ sites without vacancies, and $\langle \cdot \rangle_{\textrm{alloy}}$ denotes an ensemble average over these microstates. The sum over $\vec{\nu}$ runs over all configurations obtained by replacing a single atom in a vacancy-free configuration $\vec{\sigma}$ with a vacancy, and $f(\vec{\nu})$ returns the index of the element replaced. The factor of $\frac{1}{c}$ arises in \cref{eq:vacancy_partition_function} because each vacancy-containing configuration can be derived from $c$ possible element swaps and is therefore counted $c$ times. When sufficient vacancy sources and sinks are available, the vacancy chemical potential can be set to zero and \cref{eq:vacancy_partition_function} reduces to the expressions in \cite{belak_effect_2015,lee_modeling_2026}. The semi-grand canonical partition function then takes the form:
\begin{equation}
	\label{eq:semi_grand_partition_function_simplified}
	Z(\mu_\mathrm{Va}) = \tilde{Z} (1+\xi(\mu_\mathrm{Va}))
\end{equation}
where the dependence on $\mu_\mathrm{Va}$ is indicated explicitly, while dependence on the remaining chemical potentials, temperature, and system size is suppressed for brevity. The ensemble average of the number of atoms of type $i$ is:
\begin{widetext}
	\begin{align}
		\langle N_i \rangle
		 & =
		\sum_{\vec{\sigma}^\prime}
		N_i(\vec{\sigma}^\prime)
		\frac{\exp\left(-\beta \Omega(\vec{\sigma}^\prime)\right)}{Z}
		\nonumber \\
		 & =
		\sum_{\vec{\sigma}} \frac{\exp\left(-\beta \Omega(\vec{\sigma})\right)}{Z}
		\Bigg[
			N_i(\vec{\sigma})+
			\frac{1}{c}\sum_{\vec{\nu}}
			N_i(\vec{\nu})
			\exp\!\left(-\beta \Delta \Omega(\vec{\nu})\right)
			\Bigg]
        \label{eqn:mean_term_in_text} 
		\\
		 & = \frac{\langle N_{i} + N_{i} Z_\mathrm{Va}(\mu_\mathrm{Va})\rangle_{\textrm{alloy}} - \langle Z_\mathrm{Va}(\mu_\mathrm{Va})\rangle_{\textrm{alloy},i}}{1+\xi(\mu_\mathrm{Va})} \label{eqn:mean_term}            \\
		 & = \frac{\langle N_{i} + N_{i} \exp{\left(\beta \mu_\mathrm{Va}\right)} Z_\mathrm{Va}\rangle_{\textrm{alloy}} - \exp{\left(\beta \mu_\mathrm{Va}\right)}\langle Z_\mathrm{Va}\rangle_{\textrm{alloy},i}}{1+\exp{\left(\beta \mu_\mathrm{Va}\right)}\langle Z_\mathrm{Va}\rangle_{\textrm{alloy}} }\label{eqn:mean_term_with_muVa}
	\end{align}
\end{widetext}
where $\Delta\Omega(\vec{\nu}) = \Omega(\vec{\nu}) - \Omega(\vec{\sigma})$ and $\langle \cdot \rangle_{\textrm{alloy},i}$ denotes an ensemble average in which the sum over $\vec{\nu}$ is restricted to configurations with $f(\vec{\nu}) = i$, i.e., only atoms of species $i$ are replaced by a vacancy. Here \cref{eqn:mean_term_in_text} corresponds to \cref{eq:mean_term_general} and is expanded in \cref{eqn:mean_term,eqn:mean_term_with_muVa} to show the dependence on $\mu_\textrm{Va}$ explicitly. \Cref{eqn:mean_term} can also be used to compute the number of vacancies, $\langle N_\mathrm{Va} \rangle = \frac{\xi(\mu_\mathrm{Va})}{1+\xi(\mu_\mathrm{Va})}$. The other term required for the covariance is:
\begin{widetext}
	\begin{align}
        \label{eqn:product_term_in_text} 
		\langle N_{i}N_{j} \rangle & = 
		\sum_{\vec{\sigma}} \frac{\exp\left(-\beta \Omega(\vec{\sigma})\right)}{Z}
		\Bigg[
			N_i(\vec{\sigma})N_j(\vec{\sigma})+
			\frac{1}{c}\sum_{\vec{\nu}}
			N_i(\vec{\nu})N_j(\vec{\nu})
			\exp\!\left(-\beta \Delta \Omega(\vec{\nu})\right)
			\Bigg]
        \\
        &= 
        \frac{1}{1+\xi(\mu_\mathrm{Va})}\Bigg[\langle N_{i}N_{j} + N_{i}N_{j} Z_\mathrm{Va}(\mu_\mathrm{Va})\rangle_{\textrm{alloy}} - \langle N_{i} Z_\mathrm{Va}(\mu_\mathrm{Va})\rangle_{\textrm{alloy},j} - \langle N_{j} Z_\mathrm{Va}(\mu_\mathrm{Va})\rangle_{\textrm{alloy},i} + \delta_{ij}\langle Z_\mathrm{Va}(\mu_\mathrm{Va}) \rangle_{\textrm{alloy},i} \Bigg]
		\label{eqn:product_term_with_muVa}
        \\
	\end{align}
\end{widetext}

Computing the ensemble averages of \cref{eqn:product_term_with_muVa,eqn:mean_term_with_muVa} requires the absolute chemical potentials $\mu_{i}$ for each element in the vacancy-free alloy. These can be obtained by computing the free energy at a fixed composition through free energy integration, determining the exchange chemical potentials via Widom-type particle exchange, and extracting the absolute chemical potentials from the intercepts of a hyperplane whose slopes are the exchange chemical potentials, anchored at the computed free energy \cite{lee_modeling_2026}. In the dilute-vacancy limit, the chemical potentials of the individual elements are assumed to be unchanged by the presence of vacancies.

When sufficient vacancy sinks and sources are present, the vacancy chemical potential can be set to zero. \Cref{eqn:mean_term_with_muVa} then evaluates to:
\begin{equation}
	\label{eq:mean_term_vacancy_equilibrium}
	\langle N_i \rangle = \frac{\langle N_{i} + N_{i} Z_\mathrm{Va}\rangle_{\textrm{alloy}} - \langle Z_\mathrm{Va}\rangle_{\textrm{alloy},i}}{1+\langle Z_\mathrm{Va}\rangle_{\textrm{alloy}} }
\end{equation}
and a corresponding expression can be obtained for \cref{eqn:product_term_with_muVa} by setting $\mu_\mathrm{Va}=0$:
\begin{widetext}
	\begin{align}
		\langle N_i N_j \rangle
		 & = \frac{1}{1+\langle Z_\mathrm{Va}\rangle_{\textrm{alloy}}}\Bigg[\langle N_{i}N_{j} + N_{i}N_{j} Z_\mathrm{Va}\rangle_{\textrm{alloy}} - \langle N_{i} Z_\mathrm{Va}\rangle_{\textrm{alloy},j}- \langle N_{j} Z_\mathrm{Va}\rangle_{\textrm{alloy},i} + \delta_{ij}\langle Z_\mathrm{Va} \rangle_{\textrm{alloy},i} \Bigg]
		\label{eq:product_term_vacancy_equilibrium}
	\end{align}
\end{widetext}

A single Monte Carlo simulation of the vacancy-free alloy is sufficient to obtain ensemble averages at non-zero vacancy chemical potentials. The terms in \cref{eq:mean_term_vacancy_equilibrium} can be rescaled by $\exp{\left(\beta \mu_\mathrm{Va}\right)}$ to recover the ensemble averages at a different $\mu_\mathrm{Va}$, as given by \cref{eqn:mean_term_with_muVa}. This rescaling remains valid as long as the vacancy concentration is dilute and $\mu_\mathrm{Va}$ does not significantly modify the overall alloy composition or the chemical potentials of the individual elements.

The only difference between the formulations of \cite{belak_effect_2015,lee_modeling_2026} and the expressions in \cref{eq:mean_term_vacancy_equilibrium,eq:product_term_vacancy_equilibrium} is the treatment of vacancy swaps. In \cite{belak_effect_2015,lee_modeling_2026}, $Z_\mathrm{Va}$ is evaluated by replacing a single elemental species with a vacancy. Here, we perform swaps for all elemental species and average over them. In principle, each individual species swap yields the same result~\cite{li_vacancy_2024}, and averaging over all species reduces statistical noise and improves numerical stability.

\end{document}